\documentclass{emulateapj}
\usepackage{multirow}
\usepackage{natbib}
\bibstyle{aa}
\shorttitle{Hyperluminous QSO Environments}
\shortauthors{Trainor \& Steidel}

\newcommand{\super}[1]{\ensuremath{^\textrm{\scriptsize{#1}}}}
\newcommand{\sub}[1]{\ensuremath{_\textrm{\scriptsize{#1}}}}
\newcommand{\msun}{M\ensuremath{_{\odot}}}
\newcommand{\arcs}{\ensuremath{''}}
\newcommand{\arcm}{\ensuremath{'}}

\begin{document}


\title{The halo masses and galaxy environments of hyperluminous QSOs at $z\simeq2.7$ in the Keck Baryonic Structure Survey\altaffilmark{1}}

\author{Ryan F. Trainor}
\author{Charles C. Steidel}
\affil{Cahill Center for Astrophysics, MC 249-17, 1200 E California Blvd, Pasadena, CA 91125}


\altaffiltext{1}{Based on data obtained at the W.M. Keck Observatory, which is operated as a scientific partnership among the California Institute of Technology, the University of California, and NASA, and was made possible by the generous financial support of the W.M. Keck Foundation.}


\begin{abstract}

We present an analysis of the galaxy distribution surrounding 15 of the most luminous ($\gtrsim 10^{14}$ L$_{\odot}$; ${\rm M_{1450}} \simeq -30$) QSOs in the sky with $z\simeq 2.7$. Our data are drawn from the Keck Baryonic Structure Survey (KBSS), which has been optimized to examine the small-scale interplay between galaxies and the intergalactic medium (IGM) during the peak of the galaxy formation era at $z \sim 2-3$. In this work, we use the positions and spectroscopic redshifts of 1558 galaxies that lie within $\sim$3\arcm\, (4.2$h^{-1}$ comoving Mpc; cMpc) of the hyperluminous QSO (HLQSO) sightline in one of 15 independent survey fields, together with new measurements of the HLQSO systemic redshifts. By combining the spatial and redshift distributions, we measure the galaxy-HLQSO cross-correlation function, the galaxy-galaxy  autocorrelation function, and the characteristic scale of galaxy overdensities surrounding the sites of exceedingly rare, extremely rapid,  black hole accretion. On average, the HLQSOs lie within significant galaxy overdensities, characterized by a velocity dispersion $\sigma_v \simeq 200$ km s$^{-1}$ and a transverse angular scale of $\sim$25\arcs\, ($\sim$200 physical kpc). We argue that such scales are expected for small groups with log(M\sub{h}/\msun)~$\simeq$~13. The galaxy-HLQSO cross-correlation function has a best-fit correlation length $r_{0}\super{GQ} = (7.3 \pm 1.3)h^{-1}$ cMpc, while the galaxy autocorrelation measured from the spectroscopic galaxy sample in the same fields has $r_0\super{GG} = (6.0 \pm 0.5) h^{-1}$ cMpc. Based on a comparison with simulations evaluated at $z \sim 2.6$, these values imply that a typical galaxy lives in a host halo with  log(M\sub{h}/\msun)~=~11.9$\pm$0.1, while HLQSOs inhabit host halos of log(M\sub{h}/\msun)~=~12.3$\pm$0.5. In spite of the extremely large black hole masses implied by their observed luminosities [log(M\sub{BH}/\msun) $\gtrsim$ 9.7], it appears that HLQSOs do not require environments very different from their much less luminous QSO counterparts. Evidently, the exceedingly low space density of HLQSOs ($\lesssim 10^{-9}$ cMpc$^{-3}$) results from a one-in-a-million event on scales $<< 1$ Mpc, and not from being hosted by rare dark matter halos.

\end{abstract}

\section{Introduction}

The study of galaxies with supermassive black holes has become a topic of considerable interest, particularly since the discovery that properties of these black holes are strongly correlated with those of their host galaxies (e.g. \citealt{mag98,geb00,fer00}). The processes of supermassive black hole accretion and growth can produce spectacularly luminous QSOs, allowing their study over vast cosmological volumes ($0<z\lesssim7$). The details of these accretion processes, however, are concealed not only by distance, but also by our lack of knowledge concerning the duty cycle of AGN and the environments that drive and sustain their growth.

Because the brightest QSOs are extreme, ultra-luminous objects, it is often assumed that they must inhabit comparably rare environments. In particular, the rarity of these objects could arise because they require the highest mass dark matter halos, which are highly biased with respect to the overall matter distribution, or because of other, finely-tuned environmental factors that influence the availability of gas and the propensity for the black hole to accrete. As such, the masses and spatial distribution of the dark matter halos that host QSOs are of considerable interest, and detailed statistics on these quantities have become available through large-scale surveys, primarily through studies of QSO clustering using the two-point correlation function. Recent surveys have covered wide regions of the sky and large ranges of redshift e.g., the Sloan Digital Sky Survey (SDSS; \citealt{yor00,eis11}), the 2dF QSO Redshift Survey (2QZ; \citealt{cro04}), and the DEEP2 Redshift Survey \citep{dav03}.

Because these surveys include QSOs with a wide range of luminosities and redshifts, the QSO autocorrelation function has been frequently used to constrain QSO clustering out to high redshifts using the SDSS samples (e.g. \citealt{mye06,mye07a,she07,she10,ros09}), 2QZ samples (e.g. \citealt{por04,por06,cro05}), and combined 2dF-SDSS LRG and QSO survey (2SLAQ; survey description in \citealt{cro09}, clustering results in \citealt{daa08}). The results of these analyses are in broad agreement that QSOs inhabit host dark matter halos of mass log(M\sub{h}/\msun)$\sim$12.5 at redshifts $z \lesssim 3$. Due to the low space density of QSOs at all redshifts, these autocorrelation measurements have generally been confined to large scales, but complementary measurements have also been obtained: \citet{hen06} and \citet{she10} conducted surveys for close QSO pairs; the galaxy-QSO cross-correlation function was measured by \citet{ade05c}, in the DEEP2 survey at $z\sim 1$ by \citet{coi07}, and in a low redshift ($z<0.6$) SDSS QSO sample by \citet{pad09}.  These studies generally agree with the QSO autocorrelation results, and the mass scale log(M\sub{h}/\msun)$\sim$12.5 seems fairly well-established for the general population of QSOs at $z\lesssim 3$.

However, studies which divide the population of QSOs into specific subsamples reveal a more complicated picture of the dependence of QSO properties on halo mass.  Low-redshift studies display a possible relation between obscuration and host halo mass \citep{hic11}, which may be significant at higher redshifts, where the population of obscured QSOs is relatively unconstrained. \citet{she10} find that radio-loud QSOs are more strongly clustered than radio-quiet QSOs matched in redshift and optical luminosity. In addition, there is an expected dependence of QSO luminosity on host halo mass because the QSO luminosity depends on black hole mass, which in turn exhibits the aforementioned association with the mass of the host halo. In practice, however, QSO luminosities depend in detail on the availability of matter to accrete and the physical processes governing the efficiency with which this accretion occurs. Thus, it is perhaps not surprising that the clustering of QSOs shows little association with QSO luminosity in observations near $z \sim 2$ (e.g. \citealt{ade05c,cro05,daa08} and in simulations by \citealt{lid06}); however, \citet{she10} detect stronger clustering among the most luminous QSOs in their sample at $z>2.9$, and \citet{kru10} find that SDSS QSOs at $z\sim0.25$ cluster more strongly with increasing X-ray luminosity. Finally, the survey of close QSO pairs by \citet{hen06} reveals an excess at the smallest scales, which the authors attribute to dissipative interaction events that trigger QSO activity in rich environments. In short, the properties of QSOs are related to their host halo masses in a complex manner, and it is clear that other environmental factors are in play.

In this paper we study the environments of hyperluminous QSOs (HLQSOs; defined here by a luminosity log($\nu L_\nu$/L$_\odot$)$\gtrsim$14 at a rest-frame wavelength of 1450 \AA) at $2.5\lesssim z\lesssim 3$ by measuring the magnitude and scale of overdensities in the galaxy distribution at small ($\lesssim$3\arcm) projected distances using data from the Keck Baryonic Structure Survey (KBSS). This approach complements existing studies in numerous ways: targeting narrow fields allows us to study the local environments of these extremely rare HLQSOs, including the galaxies at comparable redshifts that lie far below the flux limits of the typical wide-field QSO surveys. In this way we are able to constrain the properties of the relatively unexplored environments of the highest-luminosity QSOs. Focusing on the brightest QSOs should reveal whether host halo mass plays a significant role in determining QSO properties, while sensitivity to the local environment may demonstrate whether these HLQSOs are associated with the types of environments where mergers and dissipative interaction are expected to be most common.

This paper is organized as follows: in \S\ref{sec:data} we discuss the observations used in this study; \S\ref{sec:overdensity} describes the techniques used to construct an unbiased measure of the galaxy distribution around the HLQSOs and our estimates of the magnitude and scale of the surrounding galaxy overdensities. In \S\ref{sec:corrfxn}, we describe and implement a method for estimating the small-scale galaxy-HLQSO correlation function and galaxy-galaxy autocorrelation function from our data along with the implied galaxy and HLQSO host halo masses. In \S\ref{sec:groups} we present evidence that the HLQSOs inhabit group-sized virialized structures conducive to merger events; a summary is given in \S\ref{sec:summary}. Throughout this paper, we will assume $\Omega$\sub{m}~=~0.3, $\Omega_\Lambda$~=~0.7, and $h = H_0/(100$ km s$^{-1})$. We have left all comoving length scales in terms of $h$ for ease of comparison to previous studies, but we quote physical scales, luminosities, and halo masses assuming $h=0.7$.  For further clarity, we denote comoving distance scales in units of cMpc (comoving Mpc) and physical scales as pkpc (physical kpc).

%
%
% Table 1: QSO redshifts
\begin{deluxetable*}{lccccr}
%\footnotesize
\tabletypesize{\footnotesize}
\tablecaption{HLQSO Redshifts and Corrections}
\tablewidth{0pt}
\tablehead{
\colhead{QSO} & \colhead{NIR Spectra Source\tablenotemark{a}} & \colhead{$z$\sub{new}\tablenotemark{b}} &  \colhead{$z$\sub{old}\tablenotemark{c}} & \colhead{$\Delta z$} & \colhead{$\Delta$v (km s$^{-1}$)}
}

\startdata
Q0100+13 (PHL957) &  Keck II/NIRSPEC   & $ 2.721\pm0.003 $ &   2.681   &  $-$0.040  &  $  -3214$~~~ \\
HS0105+1619         &  P200/TSPEC      & $ 2.652\pm0.003 $ &   2.640   &  $-$0.012  &  $  -983$~~~ \\
Q0142$-$10 (UM673a) &  Keck II/NIRSPEC   & $ 2.743\pm0.003 $ &   2.731   &  $-$0.012 &  $  -943$~~~ \\
Q0207$-$003 (UM402) &  P200/TSPEC      & $ 2.872\pm0.003 $ &   2.850   &  $-$0.022  &  $  -1699$~~~ \\
Q0449$-$1645          &  P200/TSPEC      & $ 2.684\pm0.003 $ &   2.600   &  $-$0.084  &  $  -6818$~~~ \\
Q0821+3107 (NVSS)   &  P200/TSPEC      & $ 2.616\pm0.003 $ &   2.624   &  $+$0.008  &      +686~~~  \\
Q1009+29 (CSO 38) &  Keck II/NIRSPEC   & $ 2.652\pm0.003 $ &   2.620   &  $-$0.032  &  $  -2620$~~~ \\
SBS1217+490       &  P200/TSPEC      & $ 2.704\pm0.003 $ &   2.698   &  $-$0.006  &  $  -484$~~~ \\
HS1442+2931       &  P200/TSPEC      & $ 2.660\pm0.003 $ &   2.638   &  $-$0.022  &  $  -1797$~~~ \\
HS1549+1919       &  Keck II/NIRSPEC   & $ 2.843\pm0.003 $ &   2.830   &  $-$0.013  &  $  -1011$~~~ \\
HS1603+3820       &  P200/TSPEC      & $ 2.551\pm0.003 $ &   2.510   &  $-$0.041  &  $  -3452$~~~ \\
Q1623+268 (KP77)\tablenotemark{d} & Keck II/NIRSPEC &  $ 2.5353\pm0.0005 $ &   2.518   &  $-$0.018  &  $  -1489$~~~ \\
HS1700+6416         &  Keck II/NIRSPEC   & $ 2.751\pm0.003 $ &   2.736   &  $-$0.015  &  $  -1220$~~~ \\
Q2206$-$199 (LBQS)  &  Keck II/NIRSPEC   & $ 2.573\pm0.003 $ &   2.558   &  $-$0.015  &  $  -1255$~~~ \\
Q2343+12 (also SDSS) & Keck II/NIRSPEC & $ 2.573\pm0.003 $ &   2.515   &  $-$0.058  &  $  -4854$~~~ 

\enddata
\tablenotetext{a}{Refers to the instrument used to measure the near-IR QSO spectra and redshift. NIRSPEC is used on the Keck II telescope, while P200 is the Palomar Hale 200-inch telescope, used with the TripleSpec instrument.}
\tablenotetext{b}{$z$\sub{new} refers to the redshift used in this analysis.}
\tablenotetext{c}{$z$\sub{old} refers to the previous published redshift value.}
\tablenotetext{d}{The redshift for Q1623+268 (KP77) is more tightly constrained because of the presence of narrow [\ion{O}{3}] lines at the presumed systemic redshift of the QSO.}

\label{table:qsos}
\end{deluxetable*}

%
%
% Table 2: Galaxy Samples
\begin{deluxetable*}{llccccccc}
%\footnotesize
\tabletypesize{\footnotesize}
\tablecaption{Galaxy Samples and HLQSO Properties}
\tablewidth{0pt}
\tablehead{
%\colhead{Field} & \colhead{$z$\sub{QSO}\tablenotemark{a}} & \colhead{L\sub{1450} (10$^{14}$L\sub{$_\odot$})\tablenotemark{b}} & \colhead{M\sub{BH} (10$^9$\msun)\tablenotemark{c}} & \colhead{N\sub{BX}} & \colhead{N\sub{MD}} & \colhead{N\sub{CDM}}& \colhead{N\sub{tot}} & \colhead{N\sub{1500}\tablenotemark{d}} \\
\colhead{Field} & \colhead{$z$\sub{QSO}\tablenotemark{a}} & \colhead{L\sub{1450}\tablenotemark{b}} & \colhead{M\sub{BH}\tablenotemark{c}} & \colhead{$N$\sub{BX}} & \colhead{$N$\sub{MD}} & \colhead{$N$\sub{CDM}}& \colhead{$N$\sub{tot}} & \colhead{$N$\sub{1500}\tablenotemark{d}} \\
 & & \colhead{(10$^{13}$L\sub{$_\odot$})} & \colhead{(10$^9$\msun)} & & & & & 
}

\startdata
Q0100+13        & ~2.721    & ~6.4 & ~2.0	& 68 & 12 & 15 &  95 &  7 \\
HS0105+1619     & ~2.652    & ~4.5 & ~1.4	& 74 &  6 & 23 & 103 &  7 \\
Q0142$-$10\tablenotemark{e}        & ~2.743    & $<$6.4~ &  $<$2.0~	& 75 & 13 & 16 & 104 &  1 \\
Q0207$-$003       & ~2.872    & ~6.1 & ~1.9	& 54 & 12 & 27 &  93 &  7 \\
Q0449$-$1645      & ~2.684    & ~4.0 & ~1.3	& 68 & 12 & 31 & 111 &  9 \\
Q0821+3107\tablenotemark{f}      & ~2.616    & ~4.1 & ~1.3	& 64 &  7 & 21 &  92 &  4 \\
Q1009+29        & ~2.652    & 10.9 & ~3.4	& 54 & 19 & 43 & 116 &  8 \\
SBS1217+490     & ~2.704    & ~5.1 & ~1.6	& 67 & 14 & 11 &  92 &  3 \\
Q1442+2931      & ~2.660    & ~4.9 & ~1.5	& 71 & 25 & 22 & 118 &  3 \\
HS1549+1919     & ~2.843    & 14.9 & ~4.6	& 54 & 14 & 39 & 107 & 23 \\
HS1603+3820	& ~2.551    & 11.0 & ~3.4	& 80 & 15 & 14 & 109 & 10 \\
Q1623$-$KP77\tablenotemark{f}	& ~2.5353   & ~3.2 & ~1.0	& 82 &  9 & 12 & 103 &  7 \\
HS1700+6416	& ~2.751    & 13.6 & ~4.3	& 69 & 16 & 16 & 101 &  6 \\
Q2206$-$199	& ~2.573    & ~4.5 & ~1.4	& 78 & 11 & 20 & 109 &  0 \\
Q2343+12	& ~2.573    & ~3.8 & ~1.2	& 71 &  9 & 25 & 105 &  6 
\enddata

\tablenotetext{a}{$z$\sub{QSO} refers to the systemic redshift of the field defined by the HLQSO (see Table \ref{table:qsos} and \S\ref{subsec:qsoz}).}
\tablenotetext{b}{L\sub{1450} refers to the estimated luminosity $\nu L_\nu$ near a rest-frame wavelength $\lambda\sub{rest}\simeq1450\AA$, extrapolated from the $g'$ and $r'$ magnitudes from the SDSS \citep{eis11} database when available, and otherwise from our own measurements. We have assumed $h=0.7$.}
\tablenotetext{c}{M\sub{BH} is the minimum black hole mass capable of producing a QSO with luminosity L\sub{1450}, assuming Eddington-limited accretion (\S\ref{subsec:bhmass}).}
\tablenotetext{d}{$N$\sub{1500}~=~N($|\delta v| < 1500$ km s$^{-1}$) is the number of galaxies in the field that have spectroscopic redshifts within 1500 km s$^{-1}$ of their corresponding HLQSO.}
\tablenotetext{e}{Q0142-10 (UM673a) is known to be gravitationally lensed \citep{sur87} and has an unknown magnification; the estimated luminosity and mass are therefore upper limits.}
\tablenotetext{f}{Q0821+3107 and Q1623$-$KP77 are the only HLQSOs in our sample with radio detections. Q0821 has a flux $f_\nu =162$ mJy at 4830 MHz \citep{lan90}; KP77 has a flux $f_\nu =6.4$ mJy at 1.4 GHz \citep{con98}.}
\label{table:fields}
\end{deluxetable*}

\section{Data}
\label{sec:data}

The data used in this study form part of the Keck Baryonic Structure Survey (KBSS; Steidel et al. 2012), a large sample (N$\sub{gal}=2298$) of high-redshift star-forming galaxies ($1.5<z<3.6$) close to the lines-of-sight  of 15 HLQSOs at redshifts 2.5$<$z$<$2.9. Because we have observed fields of differing solid angle around each of these HLQSOs, we standardized the fields for the purposes of this study by including only those galaxies within $\delta \theta \sim 3$\arcm\, (4.2$h^{-1}$ cMpc at the HLQSO redshifts) of the line-of-sight of the HLQSO in each, an area that is well-sampled for all 15 fields. This subset of the total KBSS dataset contains 1558 galaxies and comprises the entire sample used in this paper.

\subsection{HLQSO Redshifts}
\label{subsec:qsoz}

An important prerequisite to establishing the galaxy environment of the HLQSOs is an accurate measurement of the HLQSO systemic redshifts. Redshifts for QSOs in the range $2 \lesssim z\sub{QSO} \lesssim 3$ are typically measured from the peaks or centroids of broad emission lines of relatively high ionization species in the rest-frame far-UV (e.g., \ion{N}{5} $\lambda 1240$, \ion{C}{4} $\lambda 1549$, \ion{Si}{4} $\lambda 1399$, \ion{C}{3}] $\lambda 1909$). These lines are known to yield redshifts that differ significantly from systemic, and tend to be blue-shifted by several hundred to several thousand km s$^{-1}$ (see e.g. \citealt{mci99,ric02,gon08}). These velocity offsets also tend to increase with QSO luminosity, thus making the present sample of hyperluminous QSOs particularly susceptible to this issue. In view of the importance of precise redshifts to locate the HLQSO environments within the survey volume, we obtained near-IR spectra of the entire sample using NIRSPEC on the Keck II 10m telescope, TripleSpec on the Palomar 200-inch (5m) telescope, and in some cases, both (see Table \ref{table:qsos}).

Among the 15 HLQSOs in the sample, narrow forbidden lines ([\ion{O}{3}] $\lambda 5007$) were detected for only 2 of them (Q1623+268, Q2343+12), either because no such lines were present in the spectra (common at the highest luminosities), or because the HLQSO redshift was such that the strongest transitions fell in regions between the near-IR atmospheric bands. However, in all cases we were able to measure one or more hydrogen Balmer lines and the \ion{Mg}{2} $\lambda 2798$ line, which were mutually consistent and are known to be closer to the true systemic redshift than the high ionization lines in the UV \citep{mci99,ric02}. The redshifts obtained from the Balmer/\ion{Mg}{2} lines (which agree well with that given by [\ion{O}{3}] in the two cases where all were measured) were then subjected to several cross-checks, including:  the wavelength at the onset of the Lyman-$\alpha$ forest measured from the high resolution HLQSO spectra (a lower limit on the systemic redshift, but one which agrees to within $\Delta z \simeq 0.001$ of the Balmer line redshift in all but 2 cases); the redshift of narrow \ion{He}{2} $\lambda 1640$ in intermediate resolution optical spectra of the HLQSOs; and, in several cases, regions exhibiting narrow Ly$\alpha$ emission were discovered with small angular separations from the HLQSO, and we have found that such nebulae lie very close to the systemic redshift of the nearby HLQSO. In 2 cases (HS1603+3820 and Q1009+29) this last criterion led to a significant modification ($\Delta z \sim +0.01$, or $\sim 800$ km s$^{-1}$) of the redshift suggested by the near-IR spectroscopy. We adopt a HLQSO redshift uncertainty $\sigma\sub{z}$~=~270 km s$^{-1}$ (for those HLQSOs without measured [\ion{O}{3}] redshifts) based on the measured dispersion of the \ion{Mg}{2} line with respect to [\ion{O}{3}] by \citet{ric02}; the broad agreement among our many redshift criteria suggest that this is a conservative estimate of the redshift uncertainties.

Table \ref{table:qsos} summarizes the adopted redshifts for all 15 HLQSOs based on these considerations; also given (column 4; $z\sub{old}$) is the published redshift for each and the redshift and velocity error that would result from adopting the published values  ($\Delta z \equiv z_{old}-z_{new}$). As expected, all but one of the old redshifts are systematically too low (the median shift is $\sim -1500$ km s$^{-1}$, and the mean $\sim -2100$ km s$^{-1}$). Failure to account for these large velocity errors would severely compromise our measurements. As we show below, the measured $z\sub{QSO}$ values must be quite accurate given the very tight redshift-space correlation between the HLQSOs and the spectroscopically measured, continuum-selected galaxies nearby.

\subsection{Galaxy Redshifts}
\label{subsec:galz}
Galaxy redshifts were measured using low-resolution ($\sim$5\AA), rest-frame UV spectra obtained with the LRIS multi-object spectrograph on the Keck I telescope (\citealt{oke95,ste04}). Candidate galaxies were color-selected using the Lyman-break technique and were sorted as BX ($z \sim 2.2$), MD ($z \sim 2.6$), or CDM ($z \sim 3$) galaxies based on the color criteria discussed in \citet{ste03} and \citet{ade04}; the data collection and reduction procedures are described therein. All galaxies in the spectroscopic sample have $\mathcal{R} < 25.5$ [where $\mathcal{R} \equiv$~m\sub{AB}(6830\AA)], which corresponds to M\sub{AB}(1700\AA)~$\lesssim -19.9$ at $z \sim 2.7$ (about 1 magnitude fainter than M$_*$ at this redshift; see \citealt{red08}). Redshifts were determined by a combination of Ly$\alpha$ emission or absorption and far-UV interstellar (IS) absorption. Since Ly$\alpha$ emission tends to be redshifted with respect to the systemic redshift of the host galaxy, and interstellar absorption tends to be blueshifted (see e.g. \citealt{sha03,ade03,ste10}), we estimate each galaxy's systemic redshift via the method proposed in \citet{ade05d} and updated by \citet{ste10}. In this method, the average Ly$\alpha$ emission and IS absorption offsets are calculated based on the redshift of the H$\alpha$ nebular line (NIR spectroscopy is available for a subset of the galaxy sample), which traces ionized gas in star-forming regions of the galaxy, and is thus a more accurate estimate of the systemic redshift. \citet{rak11} derive similar corrections for the same galaxy sample using the expected symmetry of IGM absorption about the systemic redshift of the galaxy.

We estimate the systemic galaxy redshifts ($z\sub{gal}$) based on a combination of the above results. A more detailed discussion of our correction formulae can be found in Rudie et al. (2011; in prep.), but the formulae are reproduced below. For galaxies with NIR spectra (e.g. the H$\alpha$ line), the NIR redshift is used with no correction. For galaxies with measured Ly$\alpha$ emission but without interstellar absorption, we use the following estimate:

\begin{equation}
z\sub{gal} \equiv z\sub{Ly$\alpha$}+\frac{\Delta v\sub{Ly$\alpha$}}{c}\left(1+z\sub{Ly$\alpha$}\right)
\label{eq:zem}
\end{equation}

\noindent where $z\sub{Ly$\alpha$}$ is the redshift of the measured Ly$\alpha$ emission and $\Delta v\sub{Ly$\alpha$}=-300$ km s$^{-1}$ is the velocity shift needed to transform the Ly$\alpha$ redshift to the systemic value, $z\sub{gal}$.

For galaxies with interstellar absorption, we use an estimate based on the absorption redshift whether or not Ly$\alpha$ emission is present:

\begin{equation}
z\sub{gal} \equiv z\sub{IS}+\frac{\Delta v\sub{IS}}{c}\left(1+z\sub{IS}\right)
\label{eq:zabs}
\end{equation}

\noindent where $z\sub{IS}$ is the redshift of the measured interstellar absorption and $\Delta v\sub{Ly$\alpha$}=160$ km s$^{-1}$ is the velocity shift needed to transform the absorption redshift to the systemic value.

For galaxies with both interstellar absorption and Ly$\alpha$ emission, we verify that the corrected absorption redshift does not exceed the measured redshift of the Ly$\alpha$ line; that is, we verify that $z\sub{IS}<z\sub{gal}<z\sub{Ly$\alpha$}$, where $z\sub{gal}$ is calculated using Eq. \ref{eq:zabs} above. If this condition is not satisfied, we recompute the galaxy systemic redshift as the average of the absorption and emission redshifts:

\begin{equation}
z\sub{gal} \equiv \frac{z\sub{IS}+z\sub{Ly$\alpha$}}{2} \, .
\label{eq:zave}
\end{equation}

The residual redshift errors (calculated from the galaxies in the NIR sample) have a standard deviation $\sigma\sub{v,err}$~=~125 km s$^{-1}$, which we adopt as the uncertainty in our galaxy redshift measurements.

\begin{figure}[ht]
\center
\includegraphics[width=.5\textwidth]{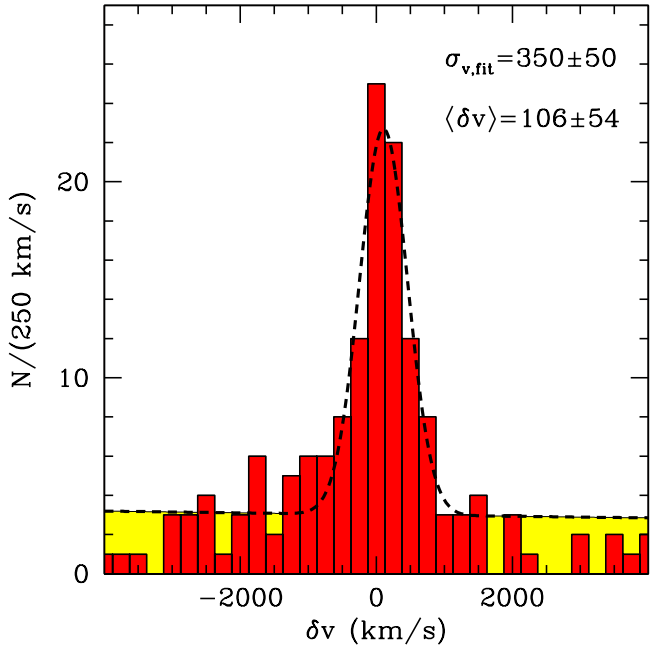}
\caption{Velocity distribution of galaxies with respect to their nearest HLQSOs, stacked for all 15 HLQSO fields. The velocity $\delta$v is given by Eq. \ref{eq:deltav}, where $\delta v=0$ for a galaxy at the redshift of its corresponding HLQSO. The yellow shaded area corresponds to the selection function, constructed as described in \S\ref{subsec:selfxn}. The dashed curve is a gaussian profile fit to the overdensity, with $\sigma\sub{v,fit}=350$ km s$^{-1}$. After removing the effect of our $\sigma\sub{v,err}\sim$ 125 km s$^{-1}$ (270 km s$^{-1}$) galaxy (HLQSO) redshift errors, we estimate a peculiar velocity scale of $\sigma\sub{v,pec}\simeq$ 200 km s$^{-1}$ for the galaxies associated with the overdensity, with an offset $\langle\delta v\rangle=$ 106$\pm$54 km s$^{-1}$ from the HLQSO redshifts.}
\label{fig:vhist}
\end{figure}

\section{Redshift Overdensity}
\label{sec:overdensity}

%Considerable care has been taken to account for selection biases and to compile a representative sample of the galaxy distribution. As our fields were observed with differing areas and numbers of observed galaxy spectra, the field dimensions were standardized to the size of our smallest fields. Any galaxies more than 4.2 $h^{-1}$ cMpc ($\sim$3\arcm) in projection from their respective central HLQSO were removed from the dataset--the numbers of galaxies in each field, as displayed in Table \ref{table:fields}, refer only to the number within these angular limits.

In order to consider the positions of the galaxies relative to their corresponding HLQSOs in redshift space while accounting for the differences in the HLQSO redshifts between fields, the redshift of each galaxy was transformed into a velocity relative to its associated HLQSO. For a galaxy with index $i$ in a field with index $j$, this velocity difference is given by

\begin{equation}
\delta v_{i,j}=\frac{c}{1+z\sub{QSO,$j$}}(z\sub{gal,$i$}-z\sub{QSO,$j$}) \,\,.
\label{eq:deltav}
\end{equation}

Once transformed to units of velocity, the distributions of galaxies relative to their HLQSOs were stacked to reveal the average environment of HLQSOs in terms of the local galaxy number density (per unit velocity)--this distribution is shown in Fig. \ref{fig:vhist}. The distribution shows a well-defined peak near $\delta v=0$, indicating the presence of significant clustering of the galaxies around the HLQSO redshifts.  We attribute the slight offset of the overdensity from the HLQSO redshifts (fit $\langle \delta v\rangle=106$$\pm$54 km s$^{-1}$) to a residual systematic offset in our determination of the HLQSO redshifts.

\begin{figure}[ht]
\center
\includegraphics[width=.45\textwidth]{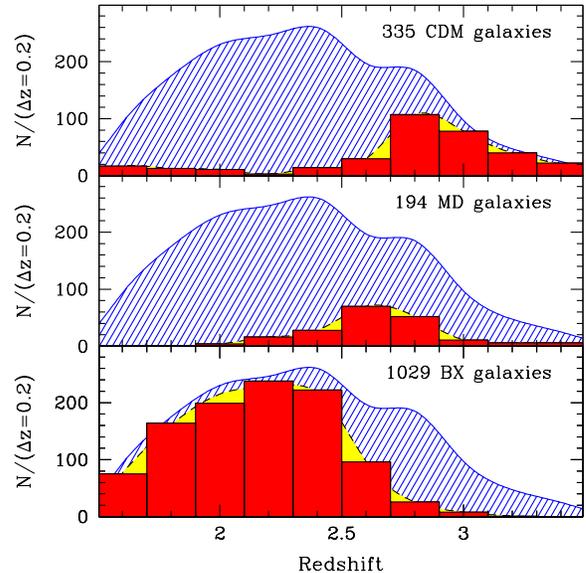}
\caption{Redshift distributions for BX, MD, and CDM color-selected galaxy types. Red histograms display the measured distributions of all such galaxies in each sample, while the yellow region represents the fit spline function specific to the color-selected sample (ie. $\mathcal{N}\sub{BX}(z)$, $\mathcal{N}\sub{MD}(z)$, and $\mathcal{N}\sub{CDM}(z)$). The blue hashed region is the overall redshift selection function for all color types.}
\label{fig:types}
\end{figure}

\subsection{Building the Selection Function}
\label{subsec:selfxn}

Clustering measurements can be grossly misinterpreted when the relevant selection functions are not well-understood \citep{ade05b}. While the criteria for selecting galaxies for follow-up spectroscopy were identical for all 15 of the KBSS fields, small differences in image depth and seeing, as well as slight changes in the algorithms used for assigning relative weights in the process of designing slit masks, can lead to field-to-field variations in the redshift selection functions. To at least partially mitigate such variations in the redshift-space sampling between fields, we used the number of successfully observed BX, MD and CDM galaxies in each field to estimate the form of our field-specific selection functions $\mathcal{N}\sub{j}(z)$. These estimates of the selection functions were constructed as follows.

First, the redshift distributions of all BX, MD and CDM galaxies in our sample were arranged in a coarse histogram with bins of width $\Delta z=0.2$. A spline fit was then performed to estimate the smooth distribution functions of each galaxy type--the histograms and spline fits for each type are displayed in Fig. \ref{fig:types}.

For each field, we built a field-specific selection function by combining these galaxy redshift distributions for each color criterion according to the number of those galaxies successfully observed in the field. Thus for a field with index $j$, the redshift selection function is given by Eq. \ref{eq:redselj}:

\begin{eqnarray}
\mathcal{N}_j(z) & = & N\sub{BX,j}\,\mathcal{N}\sub{BX}(z)+  \nonumber \\
& & N\sub{MD,j}\,\mathcal{N}\sub{MD}(z)+ \nonumber \\
& & N\sub{CDM,j}\,\mathcal{N}\sub{CDM}(z) 
\label{eq:redselj}
\end{eqnarray}

%\begin{equation}
%\mathcal{N}_j(z) = N\sub{BX,j}\,\mathcal{N}\sub{BX}(z)+N\sub{MD,j}\,\mathcal{N}\sub{MD}(z)+ N\sub{CDM,j}\,\mathcal{N}\sub{CDM}(z) 
%\label{eq:redselj}
%\end{equation}

\noindent where $N\sub{BX,j}$ corresponds to the number of BX-selected galaxies in field $j$, $\mathcal{N}\sub{BX}$ is the selection function for BX-selected galaxies over all fields, and other variables are defined similarly. We then transform these redshift-space selection functions into units of velocity relative to their corresponding bright HLQSOs using Eq. \ref{eq:deltav}. Finally, we combined this set of field-specific velocity-space selection functions (already weighted by the number of galaxies in each field) into a single stacked function:

\begin{equation}
\mathcal{N}(v)=\sum_{j=1}^{15}\mathcal{N}_j(v)  \,\,.
\label{eq:redsel}
\end{equation}

The resulting selection function is fairly flat over the range $|\delta$v$|<20000$ km s$^{-1}$ with a slight negative slope (yellow shading in Fig. \ref{fig:vhist}), indicating our slight bias toward detecting objects ``in front" of the HLQSO (that is, at lower redshifts) compared to galaxies slightly ``behind'' the HLQSO in each field. The selection function is thus a prediction for the observed distribution of galaxies in relative-velocity space in the absence of clustering.

\subsection{Bias in Field Selection}

We previously knew one KBSS field (HS1549+1919) to have a large overdensity in the galaxy distribution very close to the redshift of the central HLQSO. The variation in overdensity among fields can be estimated by $N\sub{1500}$ in Table \ref{table:fields}, which is the number of galaxies within 1500 km s$^{-1}$ of the HLQSO redshift for that field. In order to ensure that our clustering results are not being dominated by a single field, we repeated our analysis on subsamples of the data consisting of 14 of the 15 fields, removing a different field each time. In each case the magnitude and scale of the overdensity was consistent with that observed when all 15 fields were included in the analysis, indicating that the observed magnitude and scale of the overdensity are not determined by any single field.

\begin{figure}[ht]
\center
\hspace{-0.5cm}
\includegraphics[width=.35\textwidth,angle=90]{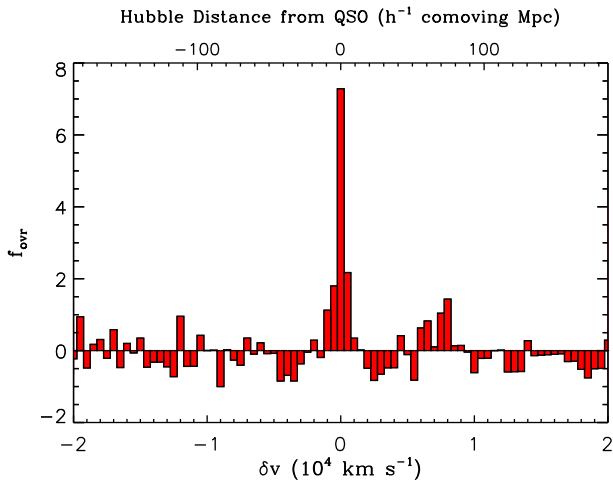}
\caption{The relative overdensity $f$\sub{ovr} (Eq. \ref{eq:fovr}) as a function of velocity relative to the central HLQSOs over a wide velocity range. The overdensity is measured in bins of 500 km s$^{-1}$, as this Hubble flow velocity roughly corresponds to the same physical scale as our transverse field of view (5$h^{-1}$ cMpc $\sim$ 500 km s$^{-1}$). See \S\ref{subsec:selfxn} for details on the selection function.}
%  The overdensities seem to exhibit some correlated structure on scales of $\sim$4000 km s$^{-1}$ (40$h^{-1}$ cMpc), discussed in \S\ref{subsec:coherent}
\label{fig:ovrden}
\end{figure}

\subsection{Redshift Clustering Results}
\label{subsec:redshiftresults}

Fig. \ref{fig:vhist} shows the observed galaxy distribution in units of velocity along with the selection function estimate from \S\ref{subsec:selfxn}. The peak in the galaxy distribution near the HLQSO redshifts is clearly visible. Fitting a Gaussian function to the histogram in Fig. \ref{fig:vhist} gives a velocity width $\sigma\sub{v,fit}=350$$\pm$$50$ km s$^{-1}$, which includes the effect our $\sigma\sub{v,err}\simeq 125$ km s$^{-1}$ galaxy redshift errors and the random residual errors in our HLQSO redshifts, assumed to be $\sigma\sub{v,err} \sim 270$ km s$^{-1}$. After subtracting the redshift errors in quadrature, we find an intrinsic velocity width of $\sigma\sub{v,pec} \simeq 200$ km s$^{-1}$ for the galaxy overdensity, which we attribute to peculiar velocities. Note that the residual HLQSO redshift errors are uncertain and likely to be largely systematic (see \S\ref{subsec:qsoz}), so our estimated velocity dispersion is an upper limit on the true peculiar velocity scale if the random component of the HLQSO redshift error is larger than we have assumed.

We also consider the relative overdensity at the HLQSO redshift by comparing the observed density to that predicted by our selection function. The distribution is plotted as a relative overdensity 

\begin{equation}
f\sub{ovr}=(N\sub{obs}-N\sub{pred})/N\sub{pred}
\label{eq:fovr}
\end{equation}

\noindent in Fig. \ref{fig:ovrden}, where $N$\sub{obs} is the number of galaxies observed in a given velocity bin and $N$\sub{pred} is the number predicted for that bin by our selection function. The relative overdensity is measured in bins with $\Delta v=500$ km s$^{-1}$; this scale was chosen to correspond roughly to the transverse scale of our field, since a Hubble-flow velocity of 500 km s$^{-1}\sim$ 5$h^{-1}$ cMpc at these redshifts. Fig. \ref{fig:ovrden} shows that the HLQSOs are associated (on average) with a $\delta n/n\sim7$ overdensity of galaxies when considered on the $\sim$5$h^{-1}$ Mpc scale of our field, with no features of comparable amplitude over a wide range of redshifts (40000 km s$^{-1}$ corresponds to $\Delta z \simeq 0.5$ at $z \simeq 2.7$).

\begin{figure}[h]
\center
\includegraphics[width=.35\textwidth,angle=90]{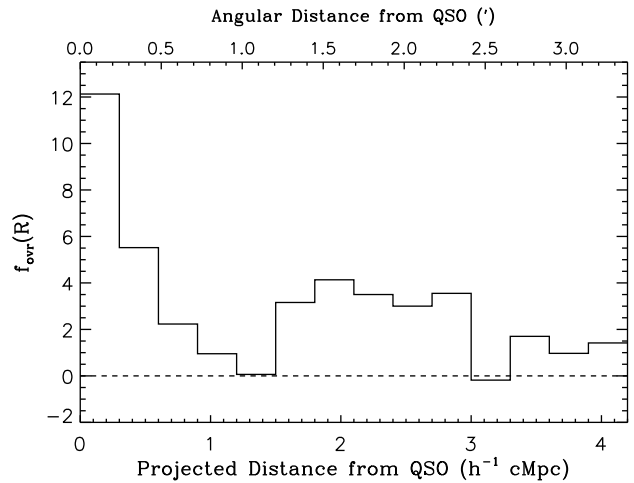}
\caption{For each projected circular annulus, the relative overdensity ($f\sub{ovr}$; Eq. \ref{eq:fovr}) of galaxies within 1500 km s$^{-1}$ of the HLQSO with respect to the redshift selection function and angular selection function. The overdensity of galaxies is localized for the most part to a tranverse scale $R\lesssim 0.5 h^{-1}$ cMpc.}
\label{fig:rdist}
\end{figure}

Repeating this analysis after dividing the galaxies into radial annuli, we find that the redshift association is most pronounced for those galaxies within 25\arcs\, of the HLQSO line of sight ($\sim$200 pkpc), though a lower level of redshift clustering does extend to larger projected distances (see Fig. \ref{fig:rdist}). If this distance is taken as an isotropic spatial scale of the galaxy overdensity, then the line-of-sight velocity dispersion due to the Hubble flow would be only $\sim$65 km s\super{-1}. However, a less-significant overdensity does extend to larger radii, and thus likely includes many galaxies that are clustered around the HLQSO but move with the Hubble flow. In order to ensure that the measured velocity width is not inflated by these non-virialized galaxies, we directly measure the velocity dispersion among the 15 galaxies within 1500 km s$^{-1}$ and 0.5 $h^{-1}$ cMpc (200 pkpc) of the HLQSOs; as discussed in \S\ref{sec:groups}, these galaxies are likely to be virialized and associated with the HLQSO, and our selection functions predict only 1.5 galaxies in this volume in the absence of clustering. These 15 galaxy velocities have a sample standard deviation of 335 km s$^{-1}$, consistent with the velocity width measured for the entire overdensity. The observed velocity spread is thus presumably set by peculiar velocities of $\sigma\sub{v,pec} \simeq 200$ km s\super{-1} among the HLQSO-associated galaxies.

A comparison of Figs. \ref{fig:ovrden} \& \ref{fig:rdist} demonstrates that the relative overdensity is highly scale-dependent.  If we assume that the width of the overdensity in velocity space is entirely due to peculiar velocities, and hence that all 15 of the galaxies observed with $R<0.5 h^{-1}$ cMpc and $|\delta v|<500$ km s$^{-1}$ are physically located within a three-dimensional distance $r<0.5 h^{-1}$ cMpc from their nearest HLQSO, then the number of galaxies in this composite volume is $\sim$50x the number predicted by our redshift and angular selection functions (described in \S\ref{subsec:selfxn} \& \S\ref{subsubsec:angselfxn}, respectively).

\section{Correlation Function Estimates}
\label{sec:corrfxn}

\subsection{Galaxy-HLQSO Cross-Correlation Function}
\label{subsec:qsocorr}
Much of the recent work on QSO clustering relies on large-scale two-point correlation functions, particularly the QSO autocorrelation function (see e.g. \citealp{she07}). The galaxy-HLQSO cross-correlation function $\xi_{Q}$ can provide a complementary estimate of HLQSO host halo mass.

The correlation function is defined as the excess conditional probability of finding a galaxy in a volume $dV$ at a distance $r=|{\bf r}_1-{\bf r}_2|$, given that there is a HLQSO at point ${\bf r}_1$, such that $P({\bf r}_2|{\bf r}_1)dV=P_0[1+\xi(r)]dV$, where $P_0 dV$ is the probability of finding a galaxy at an average place in the universe. Here we assume a power-law form for the correlation function: $\xi\sub{GQ}=(r/r_{0}\super{GQ})^{-\gamma}$, where $\gamma$ is the slope parameter and $r_0$ corresponds to the comoving distance at which the local number density of galaxies is twice that of an average place in the universe.

Many recent analyses of the two-point correlation function have dispensed with power-law fits in favor of directly modeling the halo-occupation distribution (HOD; see e.g. \citealt{sel00,ber02,zeh04}) based on the theory of \citet{pre74} and a statistical method of populating dark matter (DM) halos with galaxies. A general feature of these HOD models is a deviation from a single power law at distances near 1$h^{-1}$ cMpc due to a transition from the single-halo regime (the clustering of galaxies/QSOs within a single dark-matter halo) to the two halo regime (the clustering of galaxies/QSOs hosted by distinct halos).

In this paper we implement the simpler power-law fitting technique for the following reasons. First, our smaller sample (with respect to the large surveys at low redshift) does not allow us to detect a deviation from a power-law fit with any significance, particularly for the galaxy-HLQSO cross-correlation. Second, our choice to fix the power-law slope $\gamma$ (see below) desensitizes our result to the precise shape of the correlation function, leaving the clustering length $r_0\super{GQ}$ to primarily reflect the integrated pair-probability excess over the range of projected distances in our sample.

In practice, the three-dimensional correlation function $\xi(r)$ is not directly measurable: line-of-sight velocities are an imperfect proxy for radial distance due to peculiar velocities and redshift errors. As such, it is more useful to consider the reduced angular correlation function, $w_p(R|\Delta z)$ by integrating over a redshift or velocity window:

\begin{equation}
P(R)d\Omega=P_{0}^\prime d \Omega\left[1+w_p(R)\right]= d \Omega\int_{\Delta z} P(r) dz
\end{equation}

\noindent where $R=D_A(z) \theta (1+z)$ is the projected comoving distance from the HLQSO, and $D_A(z)$ is the angular diameter distance to the HLQSO. In the limit $\Delta z \to \infty$, and assuming a power-law form of the three-dimensional correlation function, it can be shown that the reduced angular correlation function has an equally simple power-law form:

\begin{equation}
w_p(R)=A R^{-\eta}\,\,.
\label{eq:wp}
\end{equation}
 
\noindent with $\eta=\gamma-1$. However, we would like to restrict our analysis to small redshift/velocity scales, choosing a value of $\Delta z$ that includes the entire clustering signal while eliminating the noise contribution of uncorrelated structure at large line-of-sight separations from the HLQSO, and this priority precludes the assumption of $\Delta z \to \infty$. In the case of a truncated redshift range, the reduced angular correlation function does not simplify to a power-law, and instead takes the form of a Gaussian hypergeometric function, denoted as ${}_2F_1(a, b; c; z)$. In our particular case, the reduced angular cross-correlation function $w_p\super{GQ}$ is expressed by the following:

\begin{eqnarray}
w_p\super{GQ}(R) & = & \int_{-z_0}^{z_0} (r/r_{0}\super{GQ})^{-\gamma} dz = \int_{-z_0}^{z_0} (\sqrt{R^2+z^2}/r_{0}\super{GQ})^{-\gamma} dz \nonumber \\
& = & \left(\frac{r_0\super{GQ}}{R}\right)^\gamma{}_2F_1\left(\frac{1}{2},\frac{\gamma}{2};\frac{3}{2};\frac{-z_0^2}{R^2}\right)
\label{eq:wp_model}
\end{eqnarray}

\noindent where $z_0$ is the half-width of the redshift window over which we compute the clustering strength.  We choose a value $z_0= 1500$ km s$^{-1}\simeq$ 14$h^{-1}$ cMpc in order to encompass the entire observed overdensity (see Fig. \ref{fig:vhist}) and the range of projected distances we are able to probe ($R<4.2 h^{-1}$ cMpc) without including excess noise. We then fit the reduced angular correlation function $w_p\super{GQ}(R | r_0\super{GQ}, \gamma)$ to the data by variation of the correlation length $r_0\super{GQ}$.  We fix $\gamma=1.5$ for simplicity in matching our data to halo populations (see \S\ref{subsec:halomass}); this value of $\gamma$ was chosen as it is a reasonably good fit to both the galaxy autocorrelation function and galaxy-HLQSO cross-correlation function, as well as the correlation functions among the simulated halo populations. Increasing the value of $\gamma$ causes the best-fit value of $r_0\super{GQ}$ to decrease, but the corresponding halo mass is very insensitive to the choice of $\gamma$, so long as the same value is used for both the galaxy and the simulated halo populations.

In order to estimate $w_p\super{GQ}(R)$ from our data, we separate our fields into projected circular annuli of varying widths, constructed so that each annulus has a roughly similar signal-to-noise ratio, and with our largest annulus having its outer edge $\sim$200\arcs\, from the HLQSO, a projected distance of $R=4.2h^{-1}$ cMpc. As noted above, we wish to restrict our analysis to those galaxies that are closely associated with a HLQSO in redshift as well as projected position, so we also separate our galaxy sample into two velocity groups: one with $|\delta v| \leq 1500$ km s$^{-1}$ and one with $|\delta v|>1500$ km s$^{-1}$.  In this way, we define $N_v(R_k)$ as the number of velocity-associated galaxies in the k\super{th} annular bin and $N_0(R_k)$ as the number of non-associated galaxies in the bin.\footnote[2]{In this paper, we will use the subscript or superscript $v$ to denote velocity-associated galaxies, and $0$ to denote non-associated galaxies.}  The solid angle covered by the k\super{th} annular bin is given by $\mathcal{A}_k=\pi (R\sub{outer,k}^2-R\sub{inner,k}^2)$; thus we likewise define the area densities of galaxies $\Sigma_{v,0} (R_k) = N_{v,0}(R_k)/\mathcal{A}_k$.

We then used the selection function constructed in \S\ref{subsec:selfxn} to estimate the expected number of galaxies in each velocity group [ie. $\mathcal{N}_v=\mathcal{N}(|\delta v|<1500)$ and $\mathcal{N}_0=\mathcal{N}(|\delta v|>1500)$], which we convert to expected average area densities for each velocity group ($\Sigma_{v,0}\super{pred}=\mathcal{N}_{v,0}/\mathcal{A}\sub{field}$). Finally, we divide the measured area densities by the average predicted value to define the relative overdensity of each annulus. If the overdensity $\Sigma_v(R_k)/\Sigma_v\super{pred}$ is purely due to the clustering signal, then the reduced angular cross-correlation function  is given by $w_p\super{GQ} = \Sigma\sub{v,obs}\super{GQ}(R_k)/\Sigma\sub{v,pred}\super{GQ}-1$; however, this assumption is invalid if the angular sampling of the field is not uniform, which we explore below.

\subsubsection{Angular Selection Function}
\label{subsubsec:angselfxn}

The use of the redshift selection function in $N\sub{pred}$ ensures that large-scale variations and sampling biases in redshift are taken into account in our analysis. Selection biases can also occur in the plane of the sky; because our fields are centered on their HLQSOs, any bias that varies with distance from the center of the field will mimic a change in the correlation function. To account for this effect, we recall that galaxies with $|\delta v|>1500$ km s$^{-1}$ show no association with the HLQSO (see Fig. \ref{fig:vhist}), and therefore should be uniformly distributed on average. Therefore, if the function $\Sigma_0(R)$ is not a constant, it must describe a non-uniform angular selection function, which encapsulates variations in optical selection sensitivity (e.g. due to non-uniform extinction or field coverage) as well as any biases in slit positions on our masks. We assume these biases are independent of redshift, and that they produce the same fractional excess of galaxy counts in all velocity bins. Therefore, the measured values of $\Sigma_v(R)/\Sigma_v\super{pred}$ correspond to the the true reduced correlation function $1+w_p(R)$ multiplied by a transverse (angular) selection function, which we estimate by $\Sigma_0(R_k)$.

%Our best estimate of the true correlation function is then obtained by dividing this term off our measured values of $\Sigma_v(R)$:

\begin{equation}
\frac{\Sigma_v(R_k)}{\Sigma_v\super{pred}} = \frac{\Sigma_0(R)}{\Sigma_0\super{pred}} \left[1+w_p (R_k)\right]\,\,.
\label{eq:wp_est2}
\end{equation}

%Because we limit our analysis to scales much smaller than the separations among our HLQSO fields, each galaxy is associated with only one HLQSO in the counting of pairs. As such, our transverse bins are uncorrelated--each galaxy appears in only one bin because we consider that galaxy's separation from a single HLQSO. In this case, the number of galaxy-HLQSO pairs of a given projected separation is simply the number of galaxies in the corresponding transverse bin.

We found that the measured values of $\Sigma_0(R_k)$ are well-matched by a power-law in $R$, and therefore, rather than using Eq. \ref{eq:wp_est2} directly to estimate $w_p (R_k)$, we found best-fit parameters $\alpha$ and $\beta$ for the following model:

\begin{equation}
\frac{\Sigma_0(R)}{\Sigma_0\super{pred}} = \alpha R^\beta\,\,.
\label{eq:sigma0}
\end{equation}

\begin{figure}[h]
\center
%\epsscale{0.35}
\includegraphics[width=.35\textwidth,angle=90]{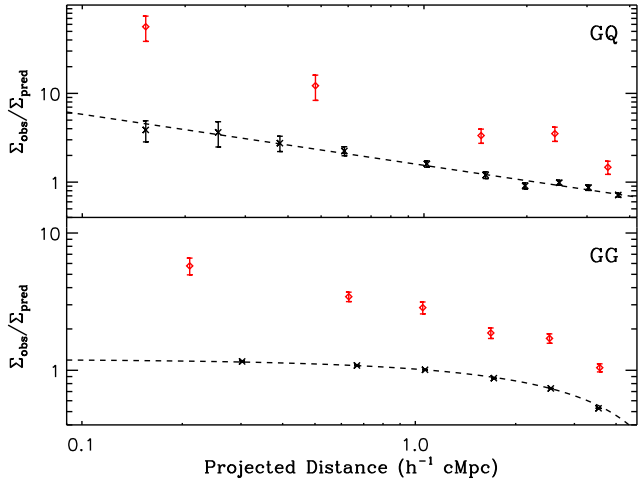}
\caption{The ratio $\Sigma\super{obs}/\Sigma\super{pred}$ for the galaxy-QSO clustering (GQ; top) and galaxy-galaxy clustering (GG; bottom). In each panel, the red diamonds denote $\Sigma_v\super{obs}/\Sigma_v\super{pred}$, while the black crosses denote $\Sigma_0\super{obs}/\Sigma_0\super{pred}$. The error bars are from Poisson uncertainties. The dashed black line is the fit to $\Sigma_0\super{obs}/\Sigma_0\super{pred}$ and defines the angular selection function; the functional form is a power-law for the GQ selection function (\S\ref{subsubsec:angselfxn}) and linear for the GG case (\S\ref{subsec:compare}).
}
\label{fig:angselfxn}
\end{figure}

The best-fit parameters for this model are $\alpha=1.59$, $\beta=-0.58$ with $R$ in $h^{-1}$ cMpc; the fit selection function is displayed in Fig. \ref{fig:angselfxn}. Combining Eqs. \ref{eq:wp_model}, \ref{eq:wp_est2}, \& \ref{eq:sigma0}, we arrive at an explicit model for $\Sigma_v(R)$ in terms of the galaxy-HLQSO cross-correlation length $r_0$ and correlation slope $\gamma$:

\begin{equation}
\frac{\Sigma_v(R_k)}{\Sigma_v\super{pred}} = \alpha R^\beta \left[1+\left(\frac{r_0}{R}\right)^\gamma {}_{2}F_1\left(\frac{1}{2},\frac{\gamma}{2};\frac{3}{2};\frac{-z_0^2}{R^2}\right)\right]
\label{eq:sigmav}
\end{equation}

\noindent where $z_0=(1500$ km s$^{-1}) H_0^{-1}(1+z)^{-1}$ is the half-width of the redshift window in physical units, and again $\alpha$ and $\beta$ are set by fitting $\Sigma_0(R)$ to $\Sigma_0(R_k)$. We then adjust the free parameter $r_0$ corresponding to the cross-correlation function to fit the measured values of $\Sigma_v(R_k)$.

The fit to $\Sigma_v$ was performed via a simple $\chi^2$-minimization using an error vector constructed assuming Poisson uncertainties in the galaxy counts. The binned data cover a range of projected distances $0.22 - 3.57 h^{-1}$ cMpc.

%This assumption is validated by the fact that our fields are well-separated on the sky with respect to the maximum distances probed by the correlation function; thus each galaxy is associated with a single HLQSO and the galaxy counts in each bin are uncorrelated with one another.

\begin{figure}[h]
\center
\includegraphics[width=.3\textwidth,angle=90]{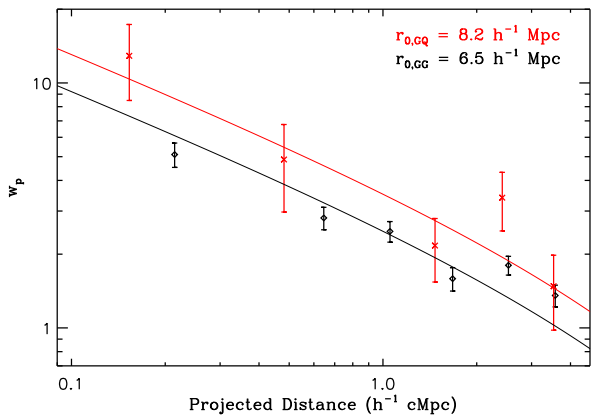}
\caption{Estimate of the reduced galaxy-HLQSO (red) and galaxy-galaxy (black) correlation functions $w_p(R)$ for those galaxies closer than 1500 km s$^{-1}$ from the HLQSO (or fiducial galaxy) redshift. $w_p$ is the excess probability of a galaxy appearing at a projected comoving separation $R$ from the HLQSO line of sight, as compared to predicted galaxy number counts determined by the redshift selection function (\S\ref{subsec:selfxn}) and angular selection function (\S\ref{subsubsec:angselfxn}). Solid curves are fits to the best-matched MultiDark halo populations (\S\ref{subsec:halomass}), which imply a galaxy halo mass log(M\sub{h,gal}/\msun)~=~11.9$\pm$0.1 (see \S\ref{subsec:millcorr}) and a HLQSO halo mass log(M\sub{h,QSO}/\msun)~=~12.3$\pm$0.5.
}
\label{fig:wp}
\end{figure}

Our empirical estimate for $w_p(R)$ obtained via the above methods is displayed in Fig. \ref{fig:wp}. We find a best-fit correlation length $r_0 = (7.3 \pm 1.3) h^{-1}$ Mpc after fixing $\gamma=1.5$, where the error is a 1$\sigma$ uncertainty computed via a bootstrap estimate. This procedure consisted of repeating the entire analysis 100 times (computation of selection functions, counting of pairs, and $\chi^2$-fitting of $w_p$) using a random bootstrap sample of 15 of the 15 independent fields selected with replacement. The quoted uncertainty is the standard deviation of parameter values derived from these 100 bootstrap samples. The results of this procedure were consistent with the results of jackknifing estimates performed using 14 of the 15 fields (ie. an $n-1$ jackknife estimate) or using 8 of the 15 fields (ie. an approximately $\sim n/2$ jackknife estimate). The $\chi^2$ value for the fit is 3.7 on 4 degrees of freedom.

As noted in \S1, \citet{ade05c} performed a cross-correlation measurement with a similar sample of color-selected galaxies to compare black hole and galaxy masses over a large range of AGN luminosities (-20 $\gtrsim$ M\sub{AB}(1350\AA) $\gtrsim$ -30) at a similar range of redshifts to our galaxy sample (1.5 $\lesssim z \lesssim$ 3.6). That study separated the AGN sample into two bins of black-hole mass, obtaining galaxy-AGN cross-correlation lengths $r_0 = 5.27_{-1.36}^{+1.59}$ $h^{-1}$ cMpc for AGNs with $10^{5.8} < M\sub{BH}/\msun < 10^8$ and $r_0 = 5.20_{-1.16}^{+1.85}$ $h^{-1}$ cMpc for AGNs with $10^{8} < M\sub{BH}/\msun < 10^{10.5}$. These measurements are fairly consistent with our own measurement of $r_0$ for the galaxy-HLQSO cross-correlation, given the size of the uncertainties, and \citet{ade05c} assume a correlation function slope of 1.6, rather than the slope of 1.5 used in this study. As noted above, the best-fit value of $r_0$ varies inversely with the chosen slope for the range of separations our measurements include, and this effect likely accounts for the slight discrepancy between these two estimates.

\subsection{Comparison to Galaxy-Galaxy Clustering}
\label{subsec:compare}
The strength of the clustering signal corresponds to the mass scale of the HLQSO-host halos, which we are interested in comparing to the average halo mass scale of non-active galaxies. As such, the relative strengths of the galaxy-galaxy (GG) and galaxy-HLQSO (GQ) clustering reveal the relative mass scales of their respective host halos, and therefore illuminate any halo-mass requirements for the formation of HLQSOs at $z\simeq 2.7$.

Our estimate of the galaxy-galaxy correlation function is based on the same technique as our galaxy-HLQSO estimates but is modified by centering on each galaxy, rather than the HLQSO, in turn. In addition, we restrict our GG analysis to those galaxies at redshifts $z\sub{gal}>2.25$ so that the GG autocorrelation function probes a similar redshift range to that of the GQ cross-correlation; the 909 galaxies with $z > 2.25$ have a median redshift $z\sub{gal}\super{med}=2.63$, while the median HLQSO redshift is $z\sub{QSO}\super{med}=2.66$. For each galaxy in our sample, we consider the number density (per unit solid angle) of galaxies as a function of projected distance from our fiducial galaxy, separating between redshift-associated galaxies (those within 1500 km s$^{-1}$ and in the same field as the fiducial galaxy) and non-associated galaxies (those outside the velocity range or in a different field). We then integrate over the redshift selection function (\S\ref{subsec:selfxn}) to find the expected number of galaxies in each interval, from which we can define an angular correlation function for each interval (by analogy to Eq. \ref{eq:wp_est2}):

\begin{equation}
\frac{\Sigma\sub{$v$,obs}\super{GG}(R)}{\Sigma\sub{$v$,pred}\super{GG}} = \frac{\Sigma\sub{0,obs}\super{GG}(R)}{\Sigma\sub{0,pred}\super{GG}}\left[1+w\sub{$p$}\super{GG} (R)\right]
\label{eq:ggeqns}
\end{equation}

\noindent where again the $v$ subscript denotes quantities corresponding to the redshift-associated sample, and the $0$ subscript denotes those corresponding to the non-associated sample. The non-associated sample is not expected to cluster about the arbitrary line of sight defined by the position of our fiducial galaxy, so we interpret the quantity $\Sigma\sub{obs,0}(R)/\Sigma\sub{pred,0}(R)$ as an estimate of the relative completeness of our angular sampling. We found the completeness (ie. the angular selection function of the galaxy-galaxy pairs) of our sample to be well-described by a linear model in $R$ with negative slope: $\Sigma_0= a R + b$. This shape reflects the fact that we are able to measure the power on small scales for essentially all galaxies, while we can see the maximum separation $\simeq 2R\sub{max}$ only for the small fraction of galaxies at the edge of our fields (and even then we see only the subset of pairs that lie entirely within the field). The best-fit parameter values for the GG angular selection function are $a=-0.160$, $b=1.04$ with $R$ in $h^{-1}$ cMpc (Fig. \ref{fig:angselfxn}).

As in the case of the GQ cross-correlation function, we then fit a model to $\Sigma_{v}\super{GG}$ that is a combination of the underlying clustering signal described by $w_{p}\super{GG}(R)$ and the selection function described by $\Sigma_0\super{GG}$. The combined model is given by Eq. \ref{eq:ggmodel}:

%\begin{eqnarray}
\begin{equation}
\frac{\Sigma_{v}\super{GG}(R)}{\Sigma_v\super{pred}} = \nonumber \\
(a R+b)\left[1+\left(\frac{r_0\super{GG}}{R}\right)^\gamma{}_2F_1\left(\frac{1}{2},\frac{\gamma}{2};\frac{3}{2};\frac{-z_0^2}{R^2}\right)\right].
\label{eq:ggmodel}
\end{equation}
%\end{eqnarray}

Fitting this model to the measured values of $\Sigma_{v}\super{GG}(R_k)$, we find the best-fit galaxy autocorrelation length to be $r_0\super{GG} = (6.0 \pm 0.5) h^{-1}$ Mpc, again fixing the slope $\gamma = 1.5$.  In this case, the errors in $\Sigma_v\super{GG}$ cannot be considered Poissonian because each galaxy is counted in several pairs and the galaxy counts are correlated between bins.  However, the 15 HLQSO fields each provide an independent estimate of $\Sigma_v\super{GG}$, and the error used for the $\chi^2$-fitting is based on the scaled scatter among these values. The quoted error on $r\sub{0}\super{GG}$ is the 1$\sigma$ uncertainty from the same bootstrap and jackknife procedures described for $r_0\super{GQ}$ (\S\ref{subsubsec:angselfxn}). The $\chi^2$ value for the fit is 9.3 on 5 degrees of freedom.

This autocorrelation length is significantly larger than that found by \citet{ade05a} [$r_0=(4.0$$\pm$$0.6)h^{-1}$ cMpc at $z=2.9$], despite both studies relying on a similarly-selected set of galaxies. However, this study is restricted to the spectroscopically-observed galaxies, which have a higher mean luminosity than the galaxies in the photometric sample used in \citet{ade05a}, and are more comparable to the higher-luminosity sub-sample of galaxies used in that paper, for which the authors estimated $r_0=(5.2$$\pm$0.6)$h^{-1}$ cMpc.  In addition, the much larger set of spectroscopic redshifts used in our sample allows us to characterize the redshift selection function with much greater accuracy, as well as to restrict our analysis to those galaxies associated with the HLQSOs in three-dimensional space.  For example, an injudicious choice of $z_0$ in Eq. \ref{eq:wp_model} would lower the estimated cross-correlation length, either by failing to count HLQSO-associated galaxies ($z_0<1500$ km s$^{-1}$) or by diluting the clustering signal by the inclusion of the voids adjacent to the HLQSO in redshift space ($z_0 >1500$ km s$^{-1}$).

\subsection{Estimate of Halo Mass}
\label{subsec:halomass}

The measured clustering of the galaxies in our sample is primarily useful in its connection to the mass scale of the galaxy host halos. Because the clustering strength of dark matter halos is a function of halo mass, we can invert this relation to obtain the halo mass for a population of objects with a given autocorrelation length. In practice, we perform this inversion numerically, finding the population of simulated halos (for which the mass is known) that match the clustering strength of our galaxy sample.

Using halo catalogs from the MultiDark MDR1 simulation (\citealt{pra11}; accessed via the MultiDark Database of \citealt{rie11}), we measured the correlation length $r_0$ as a function of minimum halo mass M\sub{h} using the Landy-Szalay estimator \citep{lan93} and assuming the same power-law slope ($\gamma=1.5$) used in our fit to the galaxy autocorrelation function. The correlation function was measured for halo populations of differing masses by varying the minimum M\sub{h} in steps of 0.05 dex; the correlation lengths for a subset of the halo samples are listed in Table \ref{table:halomass}. A power-law correlation function is a poor fit to the halo clustering at small scales due to the effect of halo exclusion; therefore, we restricted our fit to pairs with separations $1\leq d/(h^{-1}$ Mpc$)\leq 5$, a range that avoids the halo-exclusion zone while still closely matching the range of projected distances in our observed sample. In this manner, we find that our galaxy sample is most consistent with having a minimum halo mass log(M\sub{h}/\msun)$>$11.7$\pm$0.1, the fit to which is displayed in Fig. \ref{fig:wp}. The halos in this mass range have a median mass log(M\sub{h}/\msun)~=~11.9$\pm$0.1. The statistical error in the mass estimate is entirely due to the propagated error in the autocorrelation function, as the uncertainty in the autocorrelation function among simulated halos is negligible by comparison.

In addition to matching clustering strengths, we can also attempt to match the abundances of observed galaxies and simulated halos. Although our spectroscopic sample of galaxies is incomplete, we can compare this halo population to the galaxy luminosity function (GLF) of \citet{red08}, which corrects for incompleteness in both the spectroscopic and photometric samples. Luminosity functions are measured separately for galaxies with $1.9 \leq z < 2.7$ and $2.7 \leq z < 3.4$, while our sample straddles these two redshift intervals, but the GLF evolves very little over this redshift range, and the predictions of either model are quite similar.  Using the \citet{sch76} GLF parameters listed in Table 7 of \citet{red08}, and taking our magnitude limit $\mathcal{R}<25.5$ to correspond to M\sub{AB}(1700\AA)~$\lesssim -19.9$ at $z \simeq 2.7$, the Reddy et al. models predict a galaxy number density $\phi\sub{gal}=2.4-7.0\times10^{-3} h^3$Mpc$^{-3}$ (including the 1$\sigma$ limits on $\phi^*$). The number density of log(M\sub{h}/\msun)$>$11.7 halos in the MultiDark MDR1 simulation is $\phi\sub{sim}=4.4\times10^{-3} h^3$Mpc$^{-3}$, entirely consistent with the measured value of $\phi\sub{gal}$.

Taking the population of log(M\sub{h}/\msun)$>$11.7 halos to represent the host halos of the galaxies in our sample, we then estimate the mass of the HLQSO hosts by finding the population of simulated halos whose cross-correlation with the representative galaxy halos is equal to our measured galaxy-HLQSO cross-correlation function. Again varying the minimum M\sub{h} (of fiducial HLQSO hosts) in 0.05 dex increments, and using a Landy-Szalay variant for a cross-correlation with $\gamma=1.5$, we find that a HLQSO host halo mass log(M\sub{h}/\msun)$>$12.1$\pm$0.5 (a median halo mass of log(M\sub{h}/\msun)~=~12.3$\pm$0.5) is most consistent with our galaxy-HLQSO cross-correlation measurement. The error in the mass is due to the error on both $r\sub{0}\super{GQ}$ and the propagated error on the galaxy host halo mass, since the strength of the cross-correlation function depends on the mass of both the HLQSO-host and galaxy-host halo populations. The fit to the corresponding simulated cross-correlation function is shown in Fig. \ref{fig:wp}.  The MultiDark halos of log(M\sub{h}/\msun)$>$12.1$\pm$0.5 have an autocorrelation length of 6$-$15 $h^{-1}$ cMpc, which we consider to be an estimate of the HLQSO autocorrelation length, and such halos have an abundance $\phi\sub{sim}=(0.17-5.9)\times10^{-3} h^3$Mpc$^{-3}$ at z$\sim$2.5 in the simulation.

%
%
% Table 4: Simulated Halos
\begin{deluxetable}{ccc}[h]
%\footnotesize
%\tabletypesize{\scriptsize}
\tablecaption{Clustering Properties of Simulated Halos}
\tablewidth{0pt}
\tablehead{
\colhead{Minimum} & \colhead{ACF $r_0$\tablenotemark{a}} & \colhead{XCF $r_0$\tablenotemark{b}} \\
log($\frac{M\sub{h}}{\msun}$) & \colhead{($h^{-1}$ Mpc)} & \colhead{($h^{-1}$ Mpc)} 
}

\startdata
11.50 &  5.1 & 5.6 \\
11.60 &  5.6 & 5.9 \\
11.70 &  6.1 & 6.1 \\
11.80 &  6.7 & 6.3 \\
11.90 &  7.1 & 6.8 \\
12.00 &  7.8 & 6.8 \\
12.10 &  8.6 & 7.2 \\
12.20 &  9.6 & 7.6 \\
12.30 & 10.5 & 7.8 \\
12.40 & 11.6 & 7.9 \\
12.50 & 12.8 & 8.4 \\
12.60 & 14.6 & 8.6 \\
12.70 & 16.3 & 9.0 \\
12.80 & 18.6 & 9.1
\enddata

\tablenotetext{a}{Halo autocorrelation length (compare to $r_{0}\super{GG}$~=~(6.0$\pm$0.5)$h^{-1}$ cMpc)}
\tablenotetext{b}{Cross-correlation length with halos of mass log(M\sub{h}/\msun)$>$11.7 (compare to $r_{0}\super{GQ}$~=~(7.3$\pm$1.3)$h^{-1}$ cMpc)}

\label{table:halomass}
\end{deluxetable}

%\subsection{Corrections to Millennium Cosmology}
\subsection{Dependence on Simulation Cosmology}
\label{subsec:millcorr}

The MultiDark suite of simulations used cosmological parameters based on the WMAP 5-year results, \{$\Omega_m$,$\Omega_\Lambda$,$\sigma_8$,$h$\}~=~\{0.27,0.73,0.82,0.70\}, which are consistent with the most recent WMAP 7-year results from \citet{lar11}: \{0.276$\pm$0.029,0.734$\pm$0.029,0.801$\pm$0.030,0.710$\pm$0.025\}. In comparison, the older (and widely-utilized) Millennium simulation \citep{spr05} used cosmological parameters based on the WMAP 1-year results, \{$\Omega_m$,$\Omega_\Lambda$,$\sigma_8$,$h$\}~=~\{0.25,0.75,0.9,0.73\}. We here consider how such a variation in these cosmological parameters affects the halo-matching process employed in this study.

The parameters $\Omega_m$ and $\sigma_8$ both affect halo abundances, and thus affect the halo bias and the mapping from clustering strengths to halo masses. \citet{zeh11} conduct an HOD analysis on a large sample of galaxies and find that varying the matter density over the range $0.25\leq \Omega_m \leq 0.3$ produces only a $\sim$2\% variation in their clustering measurements, which is quite small compared to the statistical uncertainty in our measurements.

However, the amplitude of the linear dark-matter fluctuations, $\sigma_8$, is tied to the clustering in a more pronounced and complicated manner. The clustering of galaxies in linear theory is given by the galaxy bias and the dark-matter clustering: $\xi\sub{GG}(M) = b^2 (M)\xi\sub{DM}$. Decreasing $\sigma_8$ decreases the value of $\xi\sub{DM}$ but also greatly decreases the number density of high-mass halos, which causes the bias at a given halo mass $b(M)$ to increase.  For high-mass halos, the overall effect is to increase the clustering strength at a given mass when $\sigma_8$ is decreased, which suggests that mapping halo masses to clustering strengths using the Millennium simulation would result in a shift toward larger halo masses. Repeating our halo-matching analysis on Millennium halo catalogs, we find that the best-matched halo population of galaxies has a minimum halo mass log(M\sub{h,Mill}/\msun)$>$12.0$\pm$0.1 and a median mass log(M\sub{h,Mill}/\msun)$=$12.2$\pm$0.1 in that simulation. The discrepancy between the two simulations is $\sim$3x the statistical uncertainty in the galaxy halo mass measurements, confirming that the clustering of these massive halos is quite sensitive to the chosen cosmological parameters.

%\citet{con08} use the \citet{ade05a} galaxy sample at $z \sim 2$ and conclude that correcting the Millennium halo clustering to the WMAP 3-year parameter values (\citealt{spe07}; $\sigma_8=0.761$) lowers the estimated halo masses by 0.3$-$0.4 dex for the galaxies in that sample. As the most recent WMAP value $\sigma_8=0.801$$\pm$0.030 deviates less from the Millennium cosmology than does the value assumed by \citet{con08}, our corrections will be smaller, but we still expect that the halo masses calculated by matching to Millennium halos should be systematically lowered by a value between 0.2$-$0.3 dex to account for the updated cosmological parameter estimates.

%For this reason, {\it all the masses derived from Millennium halo matching in this paper have been lowered by 0.2 dex from the actual values in the simulation.} The quoted systematic uncertainty of 0.1 dex on these values corresponds to our imprecise knowledge of this correction.

%After this correction, we conclude that the galaxies in our sample are associated with halos of mass log(M\sub{h}\super{gal}/\msun) = 11.9$\pm$0.05$\pm$0.2 and the QSOs with halos of mass log(M\sub{h}\super{QSO}/\msun) = 11.57$\pm$0.35$\pm$0.2.

\subsection{Relative Abundances of Galaxy-Host and HLQSO-Host Halos}
\label{subsec:abundance}

The mass scales of the host halos for galaxies and HLQSOs map to halo abundances, as described in \S\ref{subsec:halomass}. A galaxy host halo abundance of $\phi\sub{sim,11.7}=4.4\times10^{-3} h^3$Mpc$^{-3}$ (the MultiDark Simulation abundance of halos with log(M\sub{h}/\msun)$>$11.7) and a HLQSO host halo abundance of $\phi\sub{sim,12.1}=1.2\times10^{-3} h^3$Mpc$^{-3}$ suggests that halos massive enough to host a HLQSO are only $\sim$4x less abundant than those massive enough to host the average galaxy in our sample; the fact that far fewer than one quarter of the galaxies in our sample host a HLQSO is a strong constraint on the duty cycle of these objects.  However, the precise value of the HLQSO duty cycle depends on the number density of HLQSOs, which in turn depends on the choice of QSO population.

All of our HLQSOs have luminosities at rest-frame 1450\AA\, of log($\nu L_\nu$/L$_\odot$)$\sim$14, or an absolute magnitude M(1450\AA)~$\sim -30$.\footnote[3]{This criterion may not be satisfied for the gravitationally lensed object Q0142-10, and it is possible that other QSOs in our sample are also lensed. However, we regard it as unlikely that significant lensing has remained undetected in these well-studied objects, so the rest-frame luminosities quoted here are assumed to be accurate.}  This is brighter than the luminosity range for which large-sample statistics are available in surveys such as SDSS and SLAQ [see e.g. \citealt{cro09}, whose M\sub{g}($z$=2) is comparable to M(1450\AA)], but we can obtain an estimate of the $z \sim 2.7$ quasar luminosity function (QLF) by extrapolating the results of the highest redshift bins of \citet{cro09} to slightly higher redshifts and luminosities; in this way we roughly estimate the number density of M(1450\AA)~$\gtrsim -30$ QSOs to be $\phi\sub{QSO} \sim 10^{-9.5}h^{3}$ Mpc$^{-3}$. Integrating this density over the total comoving volume between redshifts $2\le z \le 3$ predicts $\sim$25 QSOs in this luminosity range over the entire sky, suggesting that a large fraction of the comparably bright QSOs at these redshifts are already in our sample.

Given this number density, we can extract the duty cycle of HLQSOs from the ratio $\phi\sub{QSO}/\phi\sub{sim,12.1} \simeq 10^{-6}-10^{-7}$, defining the duty cycle as the fraction of halos massive enough to host a hyperluminous QSO [log(L/L$_\odot$)$\gtrsim$14] that actually do host such a QSO. This extreme rarity with respect to the number of potential host halos indicates that the formation of the HLQSO must rely on a correspondingly rare event occurring on scales much smaller than those probed by our analysis, perhaps related to an extremely atypical merger or galaxy interaction scenario.

\subsection{Black Hole Mass vs. Halo Mass}
\label{subsec:bhmass}

It is interesting to compare the host halo masses of the HLQSOs to the minimum black hole (BH) masses allowed by their luminosities under the assumption of Eddington-limited accretion; we will refer to this minimum mass as  M\sub{BH}. The minimum BH masses for each HLQSO are listed in Table \ref{table:fields}. We calculate the value of M\sub{BH} directly from L\sub{1450} (the value of $\nu L_\nu$ at a rest-frame wavelength of 1450\AA):

\begin{equation}
M\sub{BH} = \frac{\sigma\sub{T} L\sub{1450}}{4 \pi G m\sub{p} c} = 3.1 \times10^{-5} \left(\frac{L\sub{1450}}{L_{\odot}}\right) \msun \, .
\end{equation}

We use L\sub{1450} in place of the bolometric luminosity L\sub{bol} in order to avoid the additional uncertainty in the bolometric correction. The true bolometric correction is likely to be small: \citet{nem10} use a thin accretion disc model to predict a correction factor L\sub{bol}/L\sub{1450}~$\sim$~3, which is consistent with the empirical correction estimated by \citet{net07a} for L\sub{5100} and scaled by the L\sub{5100}/L\sub{1450} relationship of \citet{net07b}. However, there is substantial scatter in these corrections, and it may be expected that L\sub{bol}/L\sub{1450} approaches unity for QSOs selected by the most extreme rest-UV luminosities, so L\sub{1450} is a useful estimate (and likely a lower limit) on L\sub{bol}.

The HLQSOs in our sample span a range of $\sim$5x in M\sub{BH} (with the possible exception of Q0142-10), and it is notable that the field (HS1549+1919) with the largest value of M\sub{BH} is also associated with the largest redshift overdensity in the galaxy distribution (see column $N\sub{1500}$ in Table \ref{table:fields}). However, there is no clear relation between $N\sub{1500}$ and M\sub{BH} among the other fields, and our galaxy samples are not large enough to comment on the variation of M\sub{BH} with halo mass. 

The median value of M\sub{BH} for our sample is log(M\sub{BH}/\msun)~$\simeq$~9.7. This indicates that the HLQSO host DM halos are only $\sim$300-2000x more massive than their associated supermassive BHs, even assuming accretion at the Eddington limit. The relationship between BH mass and DM halo mass is uncertain even at $z\simeq 0$ (compare e.g. \citealt{fer02}, \citealt{boo10}, and \citealt{kor11}), but these BHs lie well above the predictions of the M\sub{BH}-$\sigma$ or M\sub{BH}-$v_c$ relations for any reasonable mapping of the DM halo mass to the bulge velocity dispersion $\sigma$ or circular velocity $v_c$, as demonstrated below.

Three such mappings are considered in \citet{fer02}, in which the halo virial velocity $v\sub{vir}$ is related to the circular velocity by considering $v_c=v\sub{vir}$ (a zeroth-order approximation), $v_c=1.8v\sub{vir}$ (based on observational constraints on DM halo mass profiles by \citealt{sel02}), or $v_c$/$v\sub{vir}$ given by a function of halo mass extracted from the N-body simulations of \citet{bul01}. These three different assumptions predict BH masses of log(M\sub{BH}/\msun)~=~7.0, 8.4, and 7.5, respectively for a halo of mass log(M\sub{h}/\msun)~=~12.3, corresponding to M\sub{DM}/M\sub{BH}~$\simeq$~$2\times10^5$, $8\times10^3$, and $6\times10^4$. In any of these cases, the minimum BH masses for the HLQSOs in our sample are 1-2 orders of magnitude higher than the predictions of the low-redshift associations, implying that the host halos must ``catch up'' with the BHs in order to fall on the established relations by $z \simeq 0$.

Though estimates of BH masses at high redshift are highly uncertain, as are the stellar masses of their host galaxies, this result agrees qualitatively with several observational studies that find BH host galaxies at high redshift ($1\lesssim z \lesssim 4$) of a given stellar mass have systematically higher BH masses than in the local universe (e.g. \citealt{pen06,dec10,mer10,gre10}); \citet{boo10,boo11} describe an interpretation of this evolution in terms of the compactness of DM halos, which are more tightly bound at high redshifts.  These studies generally find a smaller deviation from the $z \simeq 0$ relations than is present in our sample, but the extreme luminosities of the HLQSOs in our sample force us to probe the highest-mass end of the BH mass distribution, so our measurements are very sensitive to the scatter in the M\sub{h}$-$M\sub{BH} relation as well as evolution in the mean.

If we consider the possibility that the dynamical mass discussed in \S\ref{sec:groups} includes matter external to the HLQSO host halo at $z\simeq 2.7$, which may merge into a single more massive halo by $z \simeq 0$, we can calculate where such a halo would fall in the M\sub{h}$-$M\sub{BH} relations of \citet{fer02}. Taking a halo mass log(M\sub{h}/\msun)~=~13, the above prescriptions predict BH masses of log(M\sub{BH}/\msun)~=~8.3, 9.6 and 8.7, respectively, which lie much closer to the range of BH masses seen in our sample. However, it seems clear that the extremely high BH masses indicate that the HLQSOs are atypical (with respect to the general population of QSOs) at the smallest scales.

\section{Group-Sized HLQSO Environments}
\label{sec:groups}

In addition to the properties of the HLQSO host halos themselves, it is interesting to consider the type of larger environment these hyperluminous objects inhabit. The spatial scale of the galaxy overdensities occupied by the HLQSOs in our sample is $\sim$$0.5 h^{-1}$ cMpc (Fig. \ref{fig:rdist}; $\sim$$200$ pkpc), and the peculiar velocity scale of the composite overdensity is $\sigma\sub{v,pec}\simeq200$ km s$^{-1}$ after accounting for our measurement errors (Fig. \ref{fig:vhist}). The relatively compact nature of the overdensity suggests that it may represent a virialized structure (discussed below), in which case the inferred size and velocity scales can be combined to provide a crude estimate of the mass scale of the overdensity. The virial mass estimator can be expressed in terms of the 3D velocity dispersion $\langle{\bf v}^2\rangle$, a characteristic radius $R$, the gravitational constant $G$, and a constant $\alpha\sim1$ that depends on the geometry of the system:

\begin{equation}
M\sub{dyn} = \alpha R \langle{\bf v}^2\rangle / G \, .
\end{equation}

\noindent If we approximate our group as a sphere of uniform density, we have $\alpha=5/3$. We can also take $\langle{\bf v}^2\rangle=\langle v\sub{x}^2\rangle+\langle v\sub{y}^2\rangle+\langle v\sub{z}^2\rangle = 3 \sigma_v^2 \simeq 3\times(200$ km s$^{-1})^2$ and $R \simeq 200$ pkpc from the scales above, from which we find that the average HLQSO overdensity is associated with a total mass log(M\sub{dyn}/\msun)~$\simeq$~13 -- the approximate mass scale of a small galaxy group, and consistent with the HLQSO host halo mass derived from the clustering analysis in \S\ref{subsec:halomass}. 

Because of the crude nature of this estimate, we considered several checks to determine whether these overdensities are indeed consistent with virialized groups. If the galaxies around the HLQSOs are in virial equilibrium, their spatial extent should roughly match the virial radius $r_{200}$ of a log(M\sub{h}/\msun)~$\simeq$~13 halo, where $r\sub{200}=(3M\sub{grp}/800\pi\rho\sub{crit})^{1/3}$. In fact, the virial radius for this mass scale is approximately $r\sub{200}\simeq 200$ pkpc $\simeq 0.5 h^{-1}$ cMpc -- this close match to the observed overdensity scale suggests that the HLQSO-associated galaxy overdensities are indeed virialized.

Finally, we can estimate the number of galaxies associated with each HLQSO. From our smoothed selection function (\S\ref{subsec:selfxn}), we find that the approximate number density of spectroscopically-observed galaxies at $z\sim2.7$ is $\phi\sub{spec}=1.3\times10^{-3} h^3$Mpc$^{-3}$, while the number density of log(M\sub{h}/\msun)$>$11.7 halos in the MultiDark simulation is $\phi\sub{sim}=4.4\times10^{-3} h^3$Mpc$^{-3}$ (which is also the abundance of galaxies predicted by the GLF of \citealt{red08}). Under the assumption that each of these log(M\sub{h}/\msun)$>$11.7 dark matter halos host a galaxy of comparable luminosity to those in our sample (\S\ref{subsec:halomass}), this implies that our spectroscopic sample is $\sim$30\% complete. We find a total of 15 galaxies in our sample that are within $1500$ km s$^{-1}$ and $0.5 h^{-1}$ projected cMpc of a HLQSO; taking our completeness into account, we expect that there are another $\sim35$ galaxies remaning unobserved in this volume. On average, therefore, each of the 15 HLQSOs in our sample has $\sim$3 other log(M\sub{h}/\msun)~$\simeq$~12 galaxies within 200 pkpc, again suggesting a group-sized environment.

Note that we use the term ``environment'' here to connote a region that may or may not correspond to the host halo of the HLQSO. The mass we derive here is slightly larger than the average HLQSO host halo mass of log(M\sub{h,QSO}/\msun)~=~12.3$\pm$0.5 derived from our clustering analysis (\S\ref{subsec:halomass}), and the galaxies associated with the HLQSO overdensity extend to greater projected radii than the $\sim$130 pkpc virial radius of such a halo (Fig. \ref{fig:rdist}). The discrepancy in the mass estimate may be due to larger-than-assumed errors in the HLQSO redshifts, as noted in \S\ref{subsec:redshiftresults}; overestimation of the galaxy velocity dispersion would inflate the dynamical mass estimate of the system.  However, it may also be that the HLQSO host and its galaxy neighbors are subhalos within a larger structure corresponding to our measured dynamical mass.

In addition, we note that the velocity scale of 200$-$300 km s$^{-1}$ and the overdensity of galaxies in such an environment are extremely conducive to mergers and dissipative interactions among galaxies. We suggest that the results of this study are thus strong evidence that the fueling of these HLQSOs is associated with merger activity, with the caveat that our sample of HLQSOs are extreme outliers in the QSO luminosity distribution, and thus may be formed and sustained by rather different mechanisms than the average QSO at these redshifts.

\section{Summary}
\label{sec:summary}

We have used a large sample of galaxy redshifts to investigate the environments of 15 hyperluminous QSOs (HLQSOs) in the redshift range $2.5<z<2.9$. Our galaxy sample includes 1558 spectroscopic redshifts between $z=1.5-3.6$ from the KBSS--we use the galaxies far from the HLQSOs to characterize our redshift selection function in much greater detail than is possible with purely photometric samples. Furthermore, all the redshifts in our sample are projected within $\sim$$3$\arcm\, of one of the HLQSOs, which allows us to describe the HLQSO environments on sub-Mpc scales. The principal conclusions of this work are given here:

\begin{enumerate}

\item The HLQSOs are associated with a $\delta\sim 7$ overdensity in redshift when considered on scales of $\sim$5$h^{-1}$ Mpc. The overdensity has a velocity scale of $\sigma\sub{v,pec}\simeq200$ km s$^{-1}$ after subracting the effect of redshift errors, and a projected scale of $R\sim200$ pkpc. When stacked at the redshifts of the HLQSOs, the combined galaxy distribution shows no peaks of similar significance, and stacking on random galaxy redshifts shows that the HLQSOs are correlated with much more significant small-scale overdensities than the average galaxy in our sample.

\item Careful treatment of the HLQSO redshifts is essential in order to accurately determine which galaxies are associated with the HLQSOs in three-dimensional space.  When available, we used a combination of low-ionization broad lines, narrow emission lines, and the onset of the Ly-$\alpha$ forest in the HLQSO spectra themselves in conjunction with narrow Ly-$\alpha$ at small angular separations from the HLQSOs to obtain HLQSO redshifts offset by hundreds or thousands of km s$^{-1}$ from their previously published values.  The velocity scale of the observed overdensity, which is smaller than the measured offset for any one of these HLQSOs, demonstrates both the accuracy of our redshifts and the inadequacy of common techniques for estimating the redshifts of these hyperluminous objects.

\item The best-fit autocorrelation function for the subset of galaxies in our sample with $z>2.25$ ($z\sub{med}\simeq 2.63$) has a correlation length $r_{0}\super{GG}=(6.0\pm0.5) h^{-1}$ cMpc. Comparison to dark-matter halo catalogs from the MultiDark simulation suggests that the galaxies in our sample have a minimum halo mass of log(M\sub{h}/\msun)$>$11.7$\pm$0.1 and a median halo mass of log(M\sub{h,med}/\msun)~=~11.9$\pm$0.1.

\item The best-fit galaxy-HLQSO correlation function for our sample has a correlation length $r_{0}\super{GQ}=(7.3\pm1.3) h^{-1}$ cMpc. By measuring the clustering between dark matter halos of various masses and those halos having masses log(M\sub{h}/\msun)$>$11.7, we find that the cross-correlation between log(M\sub{h}/\msun)$>$11.7 halos and log(M\sub{h}/\msun)$>$12.1 halos most closely matches our observed value of $r_{0}\super{GQ}$. We therefore deduce that each HLQSO in our sample inhabits a dark matter halo with mass log(M\sub{h}/\msun)$>$12.1$\pm$0.5, which corresponds to a median halo mass of log(M\sub{h,med}/\msun)~=~12.3$\pm$0.5. The number density of these halos exceeds the number density of HLQSOs by a factor $\sim$$10^6-10^7$.

\item The HLQSO luminosities imply minimum masses log(M\sub{BH}/\msun) $\gtrsim$ 9.7, suggesting a BH-DM mass ratio M\sub{DM}/M\sub{BH} $\lesssim$ 300$-$2000 for a dark matter mass log(M\sub{DM}/msun) $\simeq$ 12.3$-$13. Such a small ratio indicates that the HLQSOs are significantly overmassive with respect to the M\sub{BH}$-$M\sub{h} relation at $z \simeq 0$, and appear overmassive with respect to equivalent relations at higher redshift (though black hole mass estimates are quite uncertain at high redshifts).

\item The HLQSOs in our sample are associated with group-sized environments with total mass log(M\sub{grp}/\msun)$\sim$13. This conclusion follows from a dynamical mass estimate from the peculiar velocities and projected scale of the galaxy overdensity, and is consistent with the virial radius and galaxy counts expected for such a group. The peculiar velocities and overdensities associated with groups strongly indicates that these HLQSOs inhabit environments where mergers and dissipative interactions are common.

\end{enumerate}

In conclusion, the results of this paper demonstrate that the host halos of HLQSOs are not rare, so the scarcity of these objects is likely due to an extremely improbable small-scale phenomenon that produces HLQSOs. Such a phenomenon could be related to an atypical galaxy interaction geometry or similar scenario: the overdense environment with small relative velocities would increase the probability of such an event, but an unusual merger configuration is likely required to generate such large black hole masses and QSO luminosities.

\vspace{7 mm}

We thank our collaborators for their important contributions to the Keck Baryonic Structure Survey over the course of many years: M. Bogosavljevic, D. Erb, D.R. Law, M. Pettini, O. Rakic, N. Reddy, G. Rudie, and A. Shapley. Thanks also to C. Bilinski for his help reducing some of the TripleSpec QSO spectra.   RFT would also like to thank B. Siana, N. Konidaris, A. Benson, C. Hirata, R. Quadri, and J.R. Gauthier for many useful discussions. We are grateful for the many useful comments we received from an anonymous referee, particularly in regard to the estimation of uncertainties in the correlation function parameters. The MultiDark Database used in this paper and the web application providing online access to it were constructed as part of the activities of the German Astrophysical Virtual Observatory as result of a collaboration between the Leibniz-Institute for Astrophysics Potsdam (AIP) and the Spanish MultiDark Consolider Project CSD2009-00064. The MultiDark simulation was run on the NASA's Pleiades supercomputer at the NASA Ames Research Center. The Millennium Simulation databases used in this paper and the web application providing online access to them were constructed as part of the activities of the German Astrophysical Virtual Observatory. We are indebted to the staff of the W.M. Keck Observatory who keep the instruments and telescopes running effectively. We also wish to extend thanks to those of Hawaiian ancestry on whose sacred mountain we are privileged to be guests. This work has been supported by the US National Science Foundation through grants AST-0606912 and AST-0908805. CCS acknowledges additional support from the John D. and Catherine T. MacArthur Foundation and the Peter and Patricia Gruber Foundation.

\end{}

\footnotesize

\bibliographystyle{apj}
\bibliography{qsoletter2010}

\begin{thebibliography}{66}
\expandafter\ifx\csname natexlab\endcsname\relax\def\natexlab#1{#1}\fi

\bibitem[{{Adelberger}(2005)}]{ade05b}
{Adelberger}, K.~L. 2005, \apj, 621, 574

\bibitem[{{Adelberger} {et~al.}(2005{\natexlab{a}}){Adelberger}, {Shapley},
  {Steidel}, {Pettini}, {Erb}, \& {Reddy}}]{ade05d}
{Adelberger}, K.~L., {Shapley}, A.~E., {Steidel}, C.~C., {Pettini}, M., {Erb},
  D.~K., \& {Reddy}, N.~A. 2005{\natexlab{a}}, \apj, 629, 636

\bibitem[{{Adelberger} \& {Steidel}(2005)}]{ade05c}
{Adelberger}, K.~L. \& {Steidel}, C.~C. 2005, \apjl, 627, L1

\bibitem[{{Adelberger} {et~al.}(2005{\natexlab{b}}){Adelberger}, {Steidel},
  {Pettini}, {Shapley}, {Reddy}, \& {Erb}}]{ade05a}
{Adelberger}, K.~L., {Steidel}, C.~C., {Pettini}, M., {Shapley}, A.~E.,
  {Reddy}, N.~A., \& {Erb}, D.~K. 2005{\natexlab{b}}, \apj, 619, 697

\bibitem[{{Adelberger} {et~al.}(2004){Adelberger}, {Steidel}, {Shapley},
  {Hunt}, {Erb}, {Reddy}, \& {Pettini}}]{ade04}
{Adelberger}, K.~L., {Steidel}, C.~C., {Shapley}, A.~E., {Hunt}, M.~P., {Erb},
  D.~K., {Reddy}, N.~A., \& {Pettini}, M. 2004, \apj, 607, 226

\bibitem[{{Adelberger} {et~al.}(2003){Adelberger}, {Steidel}, {Shapley}, \&
  {Pettini}}]{ade03}
{Adelberger}, K.~L., {Steidel}, C.~C., {Shapley}, A.~E., \& {Pettini}, M. 2003,
  \apj, 584, 45

\bibitem[{{Berlind} \& {Weinberg}(2002)}]{ber02}
{Berlind}, A.~A. \& {Weinberg}, D.~H. 2002, \apj, 575, 587

\bibitem[{{Booth} \& {Schaye}(2010)}]{boo10}
{Booth}, C.~M. \& {Schaye}, J. 2010, \mnras, 405, L1

\bibitem[{{Booth} \& {Schaye}(2011)}]{boo11}
---. 2011, \mnras, 413, 1158

\bibitem[{{Bullock} {et~al.}(2001){Bullock}, {Kolatt}, {Sigad}, {Somerville},
  {Kravtsov}, {Klypin}, {Primack}, \& {Dekel}}]{bul01}
{Bullock}, J.~S., {Kolatt}, T.~S., {Sigad}, Y., {Somerville}, R.~S.,
  {Kravtsov}, A.~V., {Klypin}, A.~A., {Primack}, J.~R., \& {Dekel}, A. 2001,
  \mnras, 321, 559

\bibitem[{{Coil} {et~al.}(2007){Coil}, {Hennawi}, {Newman}, {Cooper}, \&
  {Davis}}]{coi07}
{Coil}, A.~L., {Hennawi}, J.~F., {Newman}, J.~A., {Cooper}, M.~C., \& {Davis},
  M. 2007, \apj, 654, 115

\bibitem[{{Condon} {et~al.}(1998){Condon}, {Cotton}, {Greisen}, {Yin},
  {Perley}, {Taylor}, \& {Broderick}}]{con98}
{Condon}, J.~J., {Cotton}, W.~D., {Greisen}, E.~W., {Yin}, Q.~F., {Perley},
  R.~A., {Taylor}, G.~B., \& {Broderick}, J.~J. 1998, \aj, 115, 1693

\bibitem[{{Croom} {et~al.}(2005){Croom}, {Boyle}, {Shanks}, {Smith}, {Miller},
  {Outram}, {Loaring}, {Hoyle}, \& {da {\^A}ngela}}]{cro05}
{Croom}, S.~M., {Boyle}, B.~J., {Shanks}, T., {Smith}, R.~J., {Miller}, L.,
  {Outram}, P.~J., {Loaring}, N.~S., {Hoyle}, F., \& {da {\^A}ngela}, J. 2005,
  \mnras, 356, 415

\bibitem[{{Croom} {et~al.}(2009){Croom}, {Richards}, {Shanks}, {Boyle},
  {Strauss}, {Myers}, {Nichol}, {Pimbblet}, {Ross}, {Schneider}, {Sharp}, \&
  {Wake}}]{cro09}
{Croom}, S.~M., {Richards}, G.~T., {Shanks}, T., {Boyle}, B.~J., {Strauss},
  M.~A., {Myers}, A.~D., {Nichol}, R.~C., {Pimbblet}, K.~A., {Ross}, N.~P.,
  {Schneider}, D.~P., {Sharp}, R.~G., \& {Wake}, D.~A. 2009, \mnras, 399, 1755

\bibitem[{{Croom} {et~al.}(2004){Croom}, {Smith}, {Boyle}, {Shanks}, {Miller},
  {Outram}, \& {Loaring}}]{cro04}
{Croom}, S.~M., {Smith}, R.~J., {Boyle}, B.~J., {Shanks}, T., {Miller}, L.,
  {Outram}, P.~J., \& {Loaring}, N.~S. 2004, \mnras, 349, 1397

\bibitem[{{da {\^A}ngela} {et~al.}(2008){da {\^A}ngela}, {Shanks}, {Croom},
  {Weilbacher}, {Brunner}, {Couch}, {Miller}, {Myers}, {Nichol}, {Pimbblet},
  {de Propris}, {Richards}, {Ross}, {Schneider}, \& {Wake}}]{daa08}
{da {\^A}ngela}, J., {Shanks}, T., {Croom}, S.~M., {Weilbacher}, P., {Brunner},
  R.~J., {Couch}, W.~J., {Miller}, L., {Myers}, A.~D., {Nichol}, R.~C.,
  {Pimbblet}, K.~A., {de Propris}, R., {Richards}, G.~T., {Ross}, N.~P.,
  {Schneider}, D.~P., \& {Wake}, D. 2008, \mnras, 383, 565

\bibitem[{{Davis} {et~al.}(2003){Davis}, {Faber}, {Newman}, {Phillips},
  {Ellis}, {Steidel}, {Conselice}, {Coil}, {Finkbeiner}, {Koo}, {Guhathakurta},
  {Weiner}, {Schiavon}, {Willmer}, {Kaiser}, {Luppino}, {Wirth}, {Connolly},
  {Eisenhardt}, {Cooper}, \& {Gerke}}]{dav03}
{Davis}, M., {Faber}, S.~M., {Newman}, J., {Phillips}, A.~C., {Ellis}, R.~S.,
  {Steidel}, C.~C., {Conselice}, C., {Coil}, A.~L., {Finkbeiner}, D.~P., {Koo},
  D.~C., {Guhathakurta}, P., {Weiner}, B., {Schiavon}, R., {Willmer}, C.,
  {Kaiser}, N., {Luppino}, G.~A., {Wirth}, G., {Connolly}, A., {Eisenhardt},
  P., {Cooper}, M., \& {Gerke}, B. 2003, in Society of Photo-Optical
  Instrumentation Engineers (SPIE) Conference Series, Vol. 4834, Society of
  Photo-Optical Instrumentation Engineers (SPIE) Conference Series, ed.
  {P.~Guhathakurta}, 161--172

\bibitem[{{Decarli} {et~al.}(2010){Decarli}, {Falomo}, {Treves}, {Labita},
  {Kotilainen}, \& {Scarpa}}]{dec10}
{Decarli}, R., {Falomo}, R., {Treves}, A., {Labita}, M., {Kotilainen}, J.~K.,
  \& {Scarpa}, R. 2010, \mnras, 402, 2453

\bibitem[{{Eisenstein} {et~al.}(2011){Eisenstein}, {Weinberg}, {Agol},
  {Aihara}, {Allende Prieto}, {Anderson}, {Arns}, {Aubourg}, {Bailey},
  {Balbinot}, \& et~al.}]{eis11}
{Eisenstein}, D.~J., {Weinberg}, D.~H., {Agol}, E., {Aihara}, H., {Allende
  Prieto}, C., {Anderson}, S.~F., {Arns}, J.~A., {Aubourg}, {\'E}., {Bailey},
  S., {Balbinot}, E., \& et~al. 2011, \aj, 142, 72

\bibitem[{{Ferrarese}(2002)}]{fer02}
{Ferrarese}, L. 2002, \apj, 578, 90

\bibitem[{{Ferrarese} \& {Merritt}(2000)}]{fer00}
{Ferrarese}, L. \& {Merritt}, D. 2000, \apjl, 539, L9

\bibitem[{{Gebhardt} {et~al.}(2000){Gebhardt}, {Bender}, {Bower}, {Dressler},
  {Faber}, {Filippenko}, {Green}, {Grillmair}, {Ho}, {Kormendy}, {Lauer},
  {Magorrian}, {Pinkney}, {Richstone}, \& {Tremaine}}]{geb00}
{Gebhardt}, K., {Bender}, R., {Bower}, G., {Dressler}, A., {Faber}, S.~M.,
  {Filippenko}, A.~V., {Green}, R., {Grillmair}, C., {Ho}, L.~C., {Kormendy},
  J., {Lauer}, T.~R., {Magorrian}, J., {Pinkney}, J., {Richstone}, D., \&
  {Tremaine}, S. 2000, \apjl, 539, L13

\bibitem[{{Gon{\c c}alves} {et~al.}(2008){Gon{\c c}alves}, {Steidel}, \&
  {Pettini}}]{gon08}
{Gon{\c c}alves}, T.~S., {Steidel}, C.~C., \& {Pettini}, M. 2008, \apj, 676,
  816

\bibitem[{{Greene} {et~al.}(2010){Greene}, {Peng}, \& {Ludwig}}]{gre10}
{Greene}, J.~E., {Peng}, C.~Y., \& {Ludwig}, R.~R. 2010, \apj, 709, 937

\bibitem[{{Hennawi} {et~al.}(2006){Hennawi}, {Strauss}, {Oguri}, {Inada},
  {Richards}, {Pindor}, {Schneider}, {Becker}, {Gregg}, {Hall}, {Johnston},
  {Fan}, {Burles}, {Schlegel}, {Gunn}, {Lupton}, {Bahcall}, {Brunner}, \&
  {Brinkmann}}]{hen06}
{Hennawi}, J.~F., {Strauss}, M.~A., {Oguri}, M., {Inada}, N., {Richards},
  G.~T., {Pindor}, B., {Schneider}, D.~P., {Becker}, R.~H., {Gregg}, M.~D.,
  {Hall}, P.~B., {Johnston}, D.~E., {Fan}, X., {Burles}, S., {Schlegel}, D.~J.,
  {Gunn}, J.~E., {Lupton}, R.~H., {Bahcall}, N.~A., {Brunner}, R.~J., \&
  {Brinkmann}, J. 2006, \aj, 131, 1

\bibitem[{{Hickox} {et~al.}(2011){Hickox}, {Myers}, {Brodwin}, {Alexander},
  {Forman}, {Jones}, {Murray}, {Brown}, {Cool}, {Kochanek}, {Dey}, {Jannuzi},
  {Eisenstein}, {Assef}, {Eisenhardt}, {Gorjian}, {Stern}, {Le Floc'h},
  {Caldwell}, {Goulding}, \& {Mullaney}}]{hic11}
{Hickox}, R.~C., {Myers}, A.~D., {Brodwin}, M., {Alexander}, D.~M., {Forman},
  W.~R., {Jones}, C., {Murray}, S.~S., {Brown}, M.~J.~I., {Cool}, R.~J.,
  {Kochanek}, C.~S., {Dey}, A., {Jannuzi}, B.~T., {Eisenstein}, D., {Assef},
  R.~J., {Eisenhardt}, P.~R., {Gorjian}, V., {Stern}, D., {Le Floc'h}, E.,
  {Caldwell}, N., {Goulding}, A.~D., \& {Mullaney}, J.~R. 2011, \apj, 731, 117

\bibitem[{{Kormendy} \& {Bender}(2011)}]{kor11}
{Kormendy}, J. \& {Bender}, R. 2011, \nat, 469, 377

\bibitem[{{Krumpe} {et~al.}(2010){Krumpe}, {Miyaji}, \& {Coil}}]{kru10}
{Krumpe}, M., {Miyaji}, T., \& {Coil}, A.~L. 2010, \apj, 713, 558

\bibitem[{{Landy} \& {Szalay}(1993)}]{lan93}
{Landy}, S.~D. \& {Szalay}, A.~S. 1993, \apj, 412, 64

\bibitem[{{Langston} {et~al.}(1990){Langston}, {Heflin}, {Conner}, {Lehar},
  {Carilli}, \& {Burke}}]{lan90}
{Langston}, G.~I., {Heflin}, M.~B., {Conner}, S.~R., {Lehar}, J., {Carilli},
  C.~L., \& {Burke}, B.~F. 1990, \apjs, 72, 621

\bibitem[{{Larson} {et~al.}(2011){Larson}, {Dunkley}, {Hinshaw}, {Komatsu},
  {Nolta}, {Bennett}, {Gold}, {Halpern}, {Hill}, {Jarosik}, {Kogut}, {Limon},
  {Meyer}, {Odegard}, {Page}, {Smith}, {Spergel}, {Tucker}, {Weiland},
  {Wollack}, \& {Wright}}]{lar11}
{Larson}, D., {Dunkley}, J., {Hinshaw}, G., {Komatsu}, E., {Nolta}, M.~R.,
  {Bennett}, C.~L., {Gold}, B., {Halpern}, M., {Hill}, R.~S., {Jarosik}, N.,
  {Kogut}, A., {Limon}, M., {Meyer}, S.~S., {Odegard}, N., {Page}, L., {Smith},
  K.~M., {Spergel}, D.~N., {Tucker}, G.~S., {Weiland}, J.~L., {Wollack}, E., \&
  {Wright}, E.~L. 2011, \apjs, 192, 16

\bibitem[{{Lidz} {et~al.}(2006){Lidz}, {Hopkins}, {Cox}, {Hernquist}, \&
  {Robertson}}]{lid06}
{Lidz}, A., {Hopkins}, P.~F., {Cox}, T.~J., {Hernquist}, L., \& {Robertson}, B.
  2006, \apj, 641, 41

\bibitem[{{Magorrian} {et~al.}(1998){Magorrian}, {Tremaine}, {Richstone},
  {Bender}, {Bower}, {Dressler}, {Faber}, {Gebhardt}, {Green}, {Grillmair},
  {Kormendy}, \& {Lauer}}]{mag98}
{Magorrian}, J., {Tremaine}, S., {Richstone}, D., {Bender}, R., {Bower}, G.,
  {Dressler}, A., {Faber}, S.~M., {Gebhardt}, K., {Green}, R., {Grillmair}, C.,
  {Kormendy}, J., \& {Lauer}, T. 1998, \aj, 115, 2285

\bibitem[{{McIntosh} {et~al.}(1999){McIntosh}, {Rix}, {Rieke}, \&
  {Foltz}}]{mci99}
{McIntosh}, D.~H., {Rix}, H.-W., {Rieke}, M.~J., \& {Foltz}, C.~B. 1999, \apjl,
  517, L73

\bibitem[{{Merloni} {et~al.}(2010){Merloni}, {Bongiorno}, {Bolzonella},
  {Brusa}, {Civano}, {Comastri}, {Elvis}, {Fiore}, {Gilli}, {Hao}, {Jahnke},
  {Koekemoer}, {Lusso}, {Mainieri}, {Mignoli}, {Miyaji}, {Renzini}, {Salvato},
  {Silverman}, {Trump}, {Vignali}, {Zamorani}, {Capak}, {Lilly}, {Sanders},
  {Taniguchi}, {Bardelli}, {Carollo}, {Caputi}, {Contini}, {Coppa}, {Cucciati},
  {de la Torre}, {de Ravel}, {Franzetti}, {Garilli}, {Hasinger}, {Impey},
  {Iovino}, {Iwasawa}, {Kampczyk}, {Kneib}, {Knobel}, {Kova{\v c}},
  {Lamareille}, {Le Borgne}, {Le Brun}, {Le F{\`e}vre}, {Maier}, {Pello},
  {Peng}, {Perez Montero}, {Ricciardelli}, {Scodeggio}, {Tanaka}, {Tasca},
  {Tresse}, {Vergani}, \& {Zucca}}]{mer10}
{Merloni}, A., {Bongiorno}, A., {Bolzonella}, M., {Brusa}, M., {Civano}, F.,
  {Comastri}, A., {Elvis}, M., {Fiore}, F., {Gilli}, R., {Hao}, H., {Jahnke},
  K., {Koekemoer}, A.~M., {Lusso}, E., {Mainieri}, V., {Mignoli}, M., {Miyaji},
  T., {Renzini}, A., {Salvato}, M., {Silverman}, J., {Trump}, J., {Vignali},
  C., {Zamorani}, G., {Capak}, P., {Lilly}, S.~J., {Sanders}, D., {Taniguchi},
  Y., {Bardelli}, S., {Carollo}, C.~M., {Caputi}, K., {Contini}, T., {Coppa},
  G., {Cucciati}, O., {de la Torre}, S., {de Ravel}, L., {Franzetti}, P.,
  {Garilli}, B., {Hasinger}, G., {Impey}, C., {Iovino}, A., {Iwasawa}, K.,
  {Kampczyk}, P., {Kneib}, J.-P., {Knobel}, C., {Kova{\v c}}, K., {Lamareille},
  F., {Le Borgne}, J.-F., {Le Brun}, V., {Le F{\`e}vre}, O., {Maier}, C.,
  {Pello}, R., {Peng}, Y., {Perez Montero}, E., {Ricciardelli}, E.,
  {Scodeggio}, M., {Tanaka}, M., {Tasca}, L.~A.~M., {Tresse}, L., {Vergani},
  D., \& {Zucca}, E. 2010, \apj, 708, 137

\bibitem[{{Myers} {et~al.}(2007){Myers}, {Brunner}, {Nichol}, {Richards},
  {Schneider}, \& {Bahcall}}]{mye07a}
{Myers}, A.~D., {Brunner}, R.~J., {Nichol}, R.~C., {Richards}, G.~T.,
  {Schneider}, D.~P., \& {Bahcall}, N.~A. 2007, \apj, 658, 85

\bibitem[{{Myers} {et~al.}(2006){Myers}, {Brunner}, {Richards}, {Nichol},
  {Schneider}, {Vanden Berk}, {Scranton}, {Gray}, \& {Brinkmann}}]{mye06}
{Myers}, A.~D., {Brunner}, R.~J., {Richards}, G.~T., {Nichol}, R.~C.,
  {Schneider}, D.~P., {Vanden Berk}, D.~E., {Scranton}, R., {Gray}, A.~G., \&
  {Brinkmann}, J. 2006, \apj, 638, 622

\bibitem[{{Nemmen} \& {Brotherton}(2010)}]{nem10}
{Nemmen}, R.~S. \& {Brotherton}, M.~S. 2010, \mnras, 408, 1598

\bibitem[{{Netzer} {et~al.}(2007){Netzer}, {Lira}, {Trakhtenbrot}, {Shemmer},
  \& {Cury}}]{net07b}
{Netzer}, H., {Lira}, P., {Trakhtenbrot}, B., {Shemmer}, O., \& {Cury}, I.
  2007, \apj, 671, 1256

\bibitem[{{Netzer} \& {Trakhtenbrot}(2007)}]{net07a}
{Netzer}, H. \& {Trakhtenbrot}, B. 2007, \apj, 654, 754

\bibitem[{{Oke} {et~al.}(1995){Oke}, {Cohen}, {Carr}, {Cromer}, {Dingizian},
  {Harris}, {Labrecque}, {Lucinio}, {Schaal}, {Epps}, \& {Miller}}]{oke95}
{Oke}, J.~B., {Cohen}, J.~G., {Carr}, M., {Cromer}, J., {Dingizian}, A.,
  {Harris}, F.~H., {Labrecque}, S., {Lucinio}, R., {Schaal}, W., {Epps}, H., \&
  {Miller}, J. 1995, \pasp, 107, 375

\bibitem[{{Padmanabhan} {et~al.}(2009){Padmanabhan}, {White}, {Norberg}, \&
  {Porciani}}]{pad09}
{Padmanabhan}, N., {White}, M., {Norberg}, P., \& {Porciani}, C. 2009, \mnras,
  397, 1862

\bibitem[{{Peng} {et~al.}(2006){Peng}, {Impey}, {Rix}, {Kochanek}, {Keeton},
  {Falco}, {Leh{\'a}r}, \& {McLeod}}]{pen06}
{Peng}, C.~Y., {Impey}, C.~D., {Rix}, H.-W., {Kochanek}, C.~S., {Keeton},
  C.~R., {Falco}, E.~E., {Leh{\'a}r}, J., \& {McLeod}, B.~A. 2006, \apj, 649,
  616

\bibitem[{{Porciani} {et~al.}(2004){Porciani}, {Magliocchetti}, \&
  {Norberg}}]{por04}
{Porciani}, C., {Magliocchetti}, M., \& {Norberg}, P. 2004, \mnras, 355, 1010

\bibitem[{{Porciani} \& {Norberg}(2006)}]{por06}
{Porciani}, C. \& {Norberg}, P. 2006, \mnras, 371, 1824

\bibitem[{{Prada} {et~al.}(2011){Prada}, {Klypin}, {Cuesta}, {Betancort-Rijo},
  \& {Primack}}]{pra11}
{Prada}, F., {Klypin}, A.~A., {Cuesta}, A.~J., {Betancort-Rijo}, J.~E., \&
  {Primack}, J. 2011, ArXiv e-prints

\bibitem[{{Press} \& {Schechter}(1974)}]{pre74}
{Press}, W.~H. \& {Schechter}, P. 1974, \apj, 187, 425

\bibitem[{{Rakic} {et~al.}(2011){Rakic}, {Schaye}, {Steidel}, \&
  {Rudie}}]{rak11}
{Rakic}, O., {Schaye}, J., {Steidel}, C.~C., \& {Rudie}, G.~C. 2011, \mnras,
  414, 3265

\bibitem[{{Reddy} {et~al.}(2008){Reddy}, {Steidel}, {Pettini}, {Adelberger},
  {Shapley}, {Erb}, \& {Dickinson}}]{red08}
{Reddy}, N.~A., {Steidel}, C.~C., {Pettini}, M., {Adelberger}, K.~L.,
  {Shapley}, A.~E., {Erb}, D.~K., \& {Dickinson}, M. 2008, \apjs, 175, 48

\bibitem[{{Richards} {et~al.}(2002){Richards}, {Vanden Berk}, {Reichard},
  {Hall}, {Schneider}, {SubbaRao}, {Thakar}, \& {York}}]{ric02}
{Richards}, G.~T., {Vanden Berk}, D.~E., {Reichard}, T.~A., {Hall}, P.~B.,
  {Schneider}, D.~P., {SubbaRao}, M., {Thakar}, A.~R., \& {York}, D.~G. 2002,
  \aj, 124, 1

\bibitem[{{Riebe} {et~al.}(2011){Riebe}, {Partl}, {Enke}, {Forero-Romero},
  {Gottloeber}, {Klypin}, {Lemson}, {Prada}, {Primack}, {Steinmetz}, \&
  {Turchaninov}}]{rie11}
{Riebe}, K., {Partl}, A.~M., {Enke}, H., {Forero-Romero}, J., {Gottloeber}, S.,
  {Klypin}, A., {Lemson}, G., {Prada}, F., {Primack}, J.~R., {Steinmetz}, M.,
  \& {Turchaninov}, V. 2011, ArXiv e-prints

\bibitem[{{Ross} {et~al.}(2009){Ross}, {Shen}, {Strauss}, {Vanden Berk},
  {Connolly}, {Richards}, {Schneider}, {Weinberg}, {Hall}, {Bahcall}, \&
  {Brunner}}]{ros09}
{Ross}, N.~P., {Shen}, Y., {Strauss}, M.~A., {Vanden Berk}, D.~E., {Connolly},
  A.~J., {Richards}, G.~T., {Schneider}, D.~P., {Weinberg}, D.~H., {Hall},
  P.~B., {Bahcall}, N.~A., \& {Brunner}, R.~J. 2009, \apj, 697, 1634

\bibitem[{{Schechter}(1976)}]{sch76}
{Schechter}, P. 1976, \apj, 203, 297

\bibitem[{{Seljak}(2000)}]{sel00}
{Seljak}, U. 2000, \mnras, 318, 203

\bibitem[{{Seljak}(2002)}]{sel02}
---. 2002, \mnras, 334, 797

\bibitem[{{Shapley} {et~al.}(2003){Shapley}, {Steidel}, {Pettini}, \&
  {Adelberger}}]{sha03}
{Shapley}, A.~E., {Steidel}, C.~C., {Pettini}, M., \& {Adelberger}, K.~L. 2003,
  \apj, 588, 65

\bibitem[{{Shen} {et~al.}(2010){Shen}, {Hennawi}, {Shankar}, {Myers},
  {Strauss}, {Djorgovski}, {Fan}, {Giocoli}, {Mahabal}, {Schneider}, \&
  {Weinberg}}]{she10}
{Shen}, Y., {Hennawi}, J.~F., {Shankar}, F., {Myers}, A.~D., {Strauss}, M.~A.,
  {Djorgovski}, S.~G., {Fan}, X., {Giocoli}, C., {Mahabal}, A., {Schneider},
  D.~P., \& {Weinberg}, D.~H. 2010, \apj, 719, 1693

\bibitem[{{Shen} {et~al.}(2007){Shen}, {Strauss}, {Oguri}, {Hennawi}, {Fan},
  {Richards}, {Hall}, {Gunn}, {Schneider}, {Szalay}, {Thakar}, {Vanden Berk},
  {Anderson}, {Bahcall}, {Connolly}, \& {Knapp}}]{she07}
{Shen}, Y., {Strauss}, M.~A., {Oguri}, M., {Hennawi}, J.~F., {Fan}, X.,
  {Richards}, G.~T., {Hall}, P.~B., {Gunn}, J.~E., {Schneider}, D.~P.,
  {Szalay}, A.~S., {Thakar}, A.~R., {Vanden Berk}, D.~E., {Anderson}, S.~F.,
  {Bahcall}, N.~A., {Connolly}, A.~J., \& {Knapp}, G.~R. 2007, \aj, 133, 2222

\bibitem[{{Springel} {et~al.}(2005){Springel}, {White}, {Jenkins}, {Frenk},
  {Yoshida}, {Gao}, {Navarro}, {Thacker}, {Croton}, {Helly}, {Peacock}, {Cole},
  {Thomas}, {Couchman}, {Evrard}, {Colberg}, \& {Pearce}}]{spr05}
{Springel}, V., {White}, S.~D.~M., {Jenkins}, A., {Frenk}, C.~S., {Yoshida},
  N., {Gao}, L., {Navarro}, J., {Thacker}, R., {Croton}, D., {Helly}, J.,
  {Peacock}, J.~A., {Cole}, S., {Thomas}, P., {Couchman}, H., {Evrard}, A.,
  {Colberg}, J., \& {Pearce}, F. 2005, \nat, 435, 629

\bibitem[{{Steidel} {et~al.}(2003){Steidel}, {Adelberger}, {Shapley},
  {Pettini}, {Dickinson}, \& {Giavalisco}}]{ste03}
{Steidel}, C.~C., {Adelberger}, K.~L., {Shapley}, A.~E., {Pettini}, M.,
  {Dickinson}, M., \& {Giavalisco}, M. 2003, \apj, 592, 728

\bibitem[{{Steidel} {et~al.}(2010){Steidel}, {Erb}, {Shapley}, {Pettini},
  {Reddy}, {Bogosavljevi{\'c}}, {Rudie}, \& {Rakic}}]{ste10}
{Steidel}, C.~C., {Erb}, D.~K., {Shapley}, A.~E., {Pettini}, M., {Reddy}, N.,
  {Bogosavljevi{\'c}}, M., {Rudie}, G.~C., \& {Rakic}, O. 2010, \apj, 717, 289

\bibitem[{{Steidel} {et~al.}(2004){Steidel}, {Shapley}, {Pettini},
  {Adelberger}, {Erb}, {Reddy}, \& {Hunt}}]{ste04}
{Steidel}, C.~C., {Shapley}, A.~E., {Pettini}, M., {Adelberger}, K.~L., {Erb},
  D.~K., {Reddy}, N.~A., \& {Hunt}, M.~P. 2004, \apj, 604, 534

\bibitem[{{Surdej} {et~al.}(1987){Surdej}, {Magain}, {Swings}, {Borgeest},
  {Courvoisier}, {Kayser}, {Kellermann}, {Kuhr}, \& {Refsdal}}]{sur87}
{Surdej}, J., {Magain}, P., {Swings}, J.-P., {Borgeest}, U., {Courvoisier},
  T.~J.-L., {Kayser}, R., {Kellermann}, K.~I., {Kuhr}, H., \& {Refsdal}, S.
  1987, \nat, 329, 695

\bibitem[{{York} {et~al.}(2000){York}, {Adelman}, {Anderson}, {Anderson},
  {Annis}, {Bahcall}, {Bakken}, {Barkhouser}, {Bastian}, {Berman}, {Boroski},
  {Bracker}, {Briegel}, {Briggs}, {Brinkmann}, {Brunner}, {Burles}, {Carey},
  {Carr}, {Castander}, {Chen}, {Colestock}, {Connolly}, {Crocker}, {Csabai},
  {Czarapata}, {Davis}, {Doi}, {Dombeck}, {Eisenstein}, {Ellman}, {Elms},
  {Evans}, {Fan}, {Federwitz}, {Fiscelli}, {Friedman}, {Frieman}, {Fukugita},
  {Gillespie}, {Gunn}, {Gurbani}, {de Haas}, {Haldeman}, {Harris}, {Hayes},
  {Heckman}, {Hennessy}, {Hindsley}, {Holm}, {Holmgren}, {Huang}, {Hull},
  {Husby}, {Ichikawa}, {Ichikawa}, {Ivezi{\'c}}, {Kent}, {Kim}, {Kinney},
  {Klaene}, {Kleinman}, {Kleinman}, {Knapp}, {Korienek}, {Kron}, {Kunszt},
  {Lamb}, {Lee}, {Leger}, {Limmongkol}, {Lindenmeyer}, {Long}, {Loomis},
  {Loveday}, {Lucinio}, {Lupton}, {MacKinnon}, {Mannery}, {Mantsch}, {Margon},
  {McGehee}, {McKay}, {Meiksin}, {Merelli}, {Monet}, {Munn}, {Narayanan},
  {Nash}, {Neilsen}, {Neswold}, {Newberg}, {Nichol}, {Nicinski}, {Nonino},
  {Okada}, {Okamura}, {Ostriker}, {Owen}, {Pauls}, {Peoples}, {Peterson},
  {Petravick}, {Pier}, {Pope}, {Pordes}, {Prosapio}, {Rechenmacher}, {Quinn},
  {Richards}, {Richmond}, {Rivetta}, {Rockosi}, {Ruthmansdorfer}, {Sandford},
  {Schlegel}, {Schneider}, {Sekiguchi}, {Sergey}, {Shimasaku}, {Siegmund},
  {Smee}, {Smith}, {Snedden}, {Stone}, {Stoughton}, {Strauss}, {Stubbs},
  {SubbaRao}, {Szalay}, {Szapudi}, {Szokoly}, {Thakar}, {Tremonti}, {Tucker},
  {Uomoto}, {Vanden Berk}, {Vogeley}, {Waddell}, {Wang}, {Watanabe},
  {Weinberg}, {Yanny}, \& {Yasuda}}]{yor00}
{York}, D.~G., {Adelman}, J., {Anderson}, Jr., J.~E., {Anderson}, S.~F.,
  {Annis}, J., {Bahcall}, N.~A., {Bakken}, J.~A., {Barkhouser}, R., {Bastian},
  S., {Berman}, E., {Boroski}, W.~N., {Bracker}, S., {Briegel}, C., {Briggs},
  J.~W., {Brinkmann}, J., {Brunner}, R., {Burles}, S., {Carey}, L., {Carr},
  M.~A., {Castander}, F.~J., {Chen}, B., {Colestock}, P.~L., {Connolly}, A.~J.,
  {Crocker}, J.~H., {Csabai}, I., {Czarapata}, P.~C., {Davis}, J.~E., {Doi},
  M., {Dombeck}, T., {Eisenstein}, D., {Ellman}, N., {Elms}, B.~R., {Evans},
  M.~L., {Fan}, X., {Federwitz}, G.~R., {Fiscelli}, L., {Friedman}, S.,
  {Frieman}, J.~A., {Fukugita}, M., {Gillespie}, B., {Gunn}, J.~E., {Gurbani},
  V.~K., {de Haas}, E., {Haldeman}, M., {Harris}, F.~H., {Hayes}, J.,
  {Heckman}, T.~M., {Hennessy}, G.~S., {Hindsley}, R.~B., {Holm}, S.,
  {Holmgren}, D.~J., {Huang}, C.-h., {Hull}, C., {Husby}, D., {Ichikawa},
  S.-I., {Ichikawa}, T., {Ivezi{\'c}}, {\v Z}., {Kent}, S., {Kim}, R.~S.~J.,
  {Kinney}, E., {Klaene}, M., {Kleinman}, A.~N., {Kleinman}, S., {Knapp},
  G.~R., {Korienek}, J., {Kron}, R.~G., {Kunszt}, P.~Z., {Lamb}, D.~Q., {Lee},
  B., {Leger}, R.~F., {Limmongkol}, S., {Lindenmeyer}, C., {Long}, D.~C.,
  {Loomis}, C., {Loveday}, J., {Lucinio}, R., {Lupton}, R.~H., {MacKinnon}, B.,
  {Mannery}, E.~J., {Mantsch}, P.~M., {Margon}, B., {McGehee}, P., {McKay},
  T.~A., {Meiksin}, A., {Merelli}, A., {Monet}, D.~G., {Munn}, J.~A.,
  {Narayanan}, V.~K., {Nash}, T., {Neilsen}, E., {Neswold}, R., {Newberg},
  H.~J., {Nichol}, R.~C., {Nicinski}, T., {Nonino}, M., {Okada}, N., {Okamura},
  S., {Ostriker}, J.~P., {Owen}, R., {Pauls}, A.~G., {Peoples}, J., {Peterson},
  R.~L., {Petravick}, D., {Pier}, J.~R., {Pope}, A., {Pordes}, R., {Prosapio},
  A., {Rechenmacher}, R., {Quinn}, T.~R., {Richards}, G.~T., {Richmond}, M.~W.,
  {Rivetta}, C.~H., {Rockosi}, C.~M., {Ruthmansdorfer}, K., {Sandford}, D.,
  {Schlegel}, D.~J., {Schneider}, D.~P., {Sekiguchi}, M., {Sergey}, G.,
  {Shimasaku}, K., {Siegmund}, W.~A., {Smee}, S., {Smith}, J.~A., {Snedden},
  S., {Stone}, R., {Stoughton}, C., {Strauss}, M.~A., {Stubbs}, C., {SubbaRao},
  M., {Szalay}, A.~S., {Szapudi}, I., {Szokoly}, G.~P., {Thakar}, A.~R.,
  {Tremonti}, C., {Tucker}, D.~L., {Uomoto}, A., {Vanden Berk}, D., {Vogeley},
  M.~S., {Waddell}, P., {Wang}, S.-i., {Watanabe}, M., {Weinberg}, D.~H.,
  {Yanny}, B., \& {Yasuda}, N. 2000, \aj, 120, 1579

\bibitem[{{Zehavi} {et~al.}(2004){Zehavi}, {Weinberg}, {Zheng}, {Berlind},
  {Frieman}, {Scoccimarro}, {Sheth}, {Blanton}, {Tegmark}, {Mo}, {Bahcall},
  {Brinkmann}, {Burles}, {Csabai}, {Fukugita}, {Gunn}, {Lamb}, {Loveday},
  {Lupton}, {Meiksin}, {Munn}, {Nichol}, {Schlegel}, {Schneider}, {SubbaRao},
  {Szalay}, {Uomoto}, \& {York}}]{zeh04}
{Zehavi}, I., {Weinberg}, D.~H., {Zheng}, Z., {Berlind}, A.~A., {Frieman},
  J.~A., {Scoccimarro}, R., {Sheth}, R.~K., {Blanton}, M.~R., {Tegmark}, M.,
  {Mo}, H.~J., {Bahcall}, N.~A., {Brinkmann}, J., {Burles}, S., {Csabai}, I.,
  {Fukugita}, M., {Gunn}, J.~E., {Lamb}, D.~Q., {Loveday}, J., {Lupton}, R.~H.,
  {Meiksin}, A., {Munn}, J.~A., {Nichol}, R.~C., {Schlegel}, D., {Schneider},
  D.~P., {SubbaRao}, M., {Szalay}, A.~S., {Uomoto}, A., \& {York}, D.~G. 2004,
  \apj, 608, 16

\bibitem[{{Zehavi} {et~al.}(2011){Zehavi}, {Zheng}, {Weinberg}, {Blanton},
  {Bahcall}, {Berlind}, {Brinkmann}, {Frieman}, {Gunn}, {Lupton}, {Nichol},
  {Percival}, {Schneider}, {Skibba}, {Strauss}, {Tegmark}, \& {York}}]{zeh11}
{Zehavi}, I., {Zheng}, Z., {Weinberg}, D.~H., {Blanton}, M.~R., {Bahcall},
  N.~A., {Berlind}, A.~A., {Brinkmann}, J., {Frieman}, J.~A., {Gunn}, J.~E.,
  {Lupton}, R.~H., {Nichol}, R.~C., {Percival}, W.~J., {Schneider}, D.~P.,
  {Skibba}, R.~A., {Strauss}, M.~A., {Tegmark}, M., \& {York}, D.~G. 2011,
  \apj, 736, 59

\end{thebibliography}
\end{document}